\documentclass[fleqn,usenatbib,useAMS]{mnras}

\usepackage{gensymb}
\usepackage{xcolor}

\usepackage{graphicx, float}	
\usepackage{amsmath}	
\usepackage{multicol}        
\usepackage{bm}		
\usepackage{pdflscape}	
\usepackage{lipsum}
\usepackage{tipa}

\usepackage[T1]{fontenc}
\usepackage{ae,aecompl}

\usepackage{newtxtext,newtxmath}

\title[SALVAGE I: central gas dictates global star formation]{SDSS-ALMA Legacy Value Archival Gas Exploration (SALVAGE) -- I: global star formation is governed by central (not global) molecular gas}

\author[S. Wilkinson et al.]{Scott Wilkinson$^{1}$\thanks{Contact e-mail: \href{mailto:swilkinson@uvic.ca}{swilkinson@uvic.ca}}, Toby Brown$^{2, 1}$, Chiara Circosta$^{3, 4}$, Sara L. Ellison$^{1}$, Blake Ledger$^{1}$, Samuel D. Fielder$^{1}$ 
\\
$^{1}$Department of Physics and Astronomy, University of Victoria, Victoria, British Columbia V8P 1A1, Canada
\\
$^{2}$Herzberg Astronomy and Astrophysics Research Centre, National Research Council of Canada, 5071 West Saanich Rd, Victoria, BC, V9E 2E7, Canada
\\
$^{3}$ESA, European Space Astronomy Centre (ESAC), Camino Bajo del Castillo s/n, 28692 Villanueva de la Cañada, Madrid, Spain
\\
$^{4}$Institut de Radioastronomie Millimétrique (IRAM), 300 rue de la Piscine, 38400 Saint-Martin-d’Hères, France
}

\date{November 9, 2025}

\pubyear{2025}

\begin{document}
\label{firstpage}
\pagerange{\pageref{firstpage}--\pageref{lastpage}}
\maketitle

\begin{abstract}

Star-forming galaxies form tight relations between their stellar mass, star-formation rate, and molecular gas reservoir on global and resolved scales. On the path to quiescence, the exchange between gas and stars must inevitably be broken. Understanding the mechanisms governing star formation and quenching therefore requires observations of both the stellar and molecular gas components. To this end, we have assembled a sample of 277  galaxies ($0.02 \lesssim z \lesssim 0.25$) with semi-resolved optical and millimetre $^{12}$CO~(1--0) data, wherein the properties of the inner $\thicksim$2 kpc can be distinguished from the outer regions. This effort was made possible by the Sloan Digital Sky Survey (SDSS) catalogues and the maturing archive of the Atacama Large (sub-)Millimetre Array (ALMA). We call this dataset the SDSS-ALMA Legacy-Value Archival Gas Exploration (SALVAGE). In this work, we leverage SALVAGE to provide a semi-resolved perspective on global scaling relations and why some galaxies deviate from them. In agreement with previous work, we find that the offset of a galaxy from the global star-forming main sequence (SFMS) is driven by its inner star formation rate. With the relative inner and outer distributions of molecular gas fraction and star formation efficiency, we investigate whether the central star formation driving global changes is due to fuel availability or efficiency. We find that the position of a galaxy within the SFMS is largely due to the inner star-formation efficiency, while departure from the SFMS is driven by availability of central gas. The central few kpc are thus the most consequential region for galaxy evolution at low redshift.

\end{abstract}


\section{Introduction}
\label{Intro}

Galaxies exhibit bimodal distributions in many key properties including colour, stellar age, star formation rate, gas content, and morphology \citep{Strateva_2001, K03, Baldry04, Driver06, Wuyts2011, Bell12}. In general, blue galaxies are those with young stellar populations, with ongoing star formation arising from the reservoir of cool molecular gas \citep{Saintonge17}, arranged in a disk morphology \citep{Wuyts2011}. In contrast, red galaxies have older stellar populations, with very little ongoing star formation and limited molecular gas reservoirs arranged in an elliptical morphology. The frequency of galaxies broadly falling into two distinct categories coherently across several properties points towards two long-term stable states of galaxies: blue star-forming disks and red quiescent ellipticals. Simulations and observations agree that over billions of years, blue star-forming spirals are evolving into red quiescent ellipticals \citep[e.g.,][]{BT87,LC93,Conselice14}. 

Stars are forged from the cold molecular gas hosted in the interstellar medium (ISM) of galaxies \citep{Saintonge22, Schinnerer24}. At the end of their lives, stars return the gaseous material back to the ISM, from which new generations of stars will form \citep{Peroux20}. Through inflows and outflows of gas from the circumgalactic medium (CGM) and beyond, the baryon cycle in the ISM maintains star formation in galaxies for billions of years \citep{Putman12, Tumlinson17}. When galaxies transition from star-forming to quiescent, the baryon cycle must ultimately be disrupted. Star formation will cease if either cool molecular gas is not available to form stars or if molecular gas is present but unable to condense into stars. 

There are many proposed mechanisms that halt star formation in galaxies and evolve them to quiescence. External forces include galaxy mergers \citep{Hopkins06, Ellison22, Wilkinson22}, ram pressure stripping \citep{GunnGott72, Roberts21, Boselli22, Werle25}, and harassment \citep{Moore96, Moore98}. Internal mechanisms include stellar feedback \citep{Heckman90, MacLow99, Bolatto13-starburst}, feedback from active galactic nuclei \citep[AGN;][]{Hopkins08, King15}, and morphological quenching \citep{Martig09, Khoperskov18}. Ubiquitously, these mechanisms shut down star formation by either removing the molecular gas (e.g. ram pressure, stellar/AGN feedback), preventing the molecular gas from collapsing into stars (e.g. morphological quenching, stellar/AGN feedback), and preventing new gas from entering the galaxy and cooling enough to form stars (e.g. starvation, preventative AGN feedback, harrassment). Indeed, in many cases, there may be multiple mechanisms involved (and possibly required) to fully transition a galaxy from star-forming to quiescence \citep[e.g.][]{Trussler20}.

Differentiating between and understanding the processes that shut down star formation in galaxies requires a statistical sampling of galaxies with observations of both the stellar and gaseous components. Significant information about the stellar components of galaxies can be garnered from optical wavelengths. There are many overlapping photometric and spectroscopic surveys in the optical for this purpose such as SDSS \citep{SDSS-DR7} and UNIONS 
\citep{Gwyn25} or DESI \citep{Levi19} and DECaLS \citep{Dey19}. Although the bulk of molecular gas is in the form of molecular hydrogen, its weak rotational lines make it notoriously difficult to observe \citep{Kennicutt12}. Instead, low-J CO rotational transition emission lines have become a popular proxy due to the abundance of CO and its strong emission in the (sub-)mm \citep{Bolatto13, Schinnerer24}. 

Unresolved, global measurements of the gas and stars in galaxies are efficient for assembling large samples, but are limited to only investigating the bulk global exchange between gas and stars. Some examples include xCOLD GASS \citep{Saintonge17}, MASCOT \citep{Wylezalek22}, and ALLSMOG \citep{Bothwell14}, assembling samples of 532, 187, and 88 galaxies, respectively. These large surveys have enabled strong characterizations of global scaling relations, including the star-forming main sequence (SFMS), the Kennicutt-Schmidt relation (KS) and the molecular gas main sequence (MGMS), as well as galaxies that deviate from them \citep{Janowiecki20, Hagedorn24}.

Resolved integral field spectroscopy combined with interferometric observations in the sub-mm has allowed for a more detailed understanding of the interplay between gas and stars. For example, the SFMS, MGMS and KS relations have been found to vary within a galaxy, but hold for each individual kpc-scale region \citep{Lin19, Sanchez21, Jimenez-Donaire23}. Resolved measurements of the gas and stars also allow investigations into specific quenching mechanisms such as bars \citep{Chown19, Hogarth24}, mergers \citep{Thorp22, Garay-Solis23}, AGN feedback \citep{Ellison21, Bazzi25}, and ram pressure \citep{Zabel22, Brown23}, as well as samples of quenching galaxies such as green valley galaxies \citep{Lin22, Villanueva24} and post-starburst galaxies \citep{Otter22}. However, the detail revealed by resolved measurements comes at a cost of time-consuming observations at often oversubscribed observatories, making it challenging to assemble large statistical samplings of galaxies. Some examples of surveys seeking resolved optical and mm spectroscopy include ALMaQUEST \citep{Lin20, Ellison24-aqpde}, EDGE-CALIFA \citep{Wong24}, and VERTICO \citep{Brown21}, assembling samples of 66, 126, and 51 galaxies, respectively. The highest-resolution survey of galaxies with combined optical and millimeter data products is PHANGS-MUSE \citep{Emsellem22}, which has assembled a sample of 16 galaxies with resolution on the scale of individual star-forming clouds. 

The sample size of the combined optical and mm surveys trends with the resolution of the molecular gas observations; since lower-resolution observations are easier to obtain, it makes sense to assemble a larger sample to capture a larger variety of galaxies. The existing surveys tend to be clustered into three groups: unresolved surveys (only global values), resolved surveys (with approximately 1 kpc resolution) and the cloud-scale resolution of PHANGS (with approximately 100 pc resolution). As the resolution improves, more detail about the inner workings of galaxies is revealed. However, a smaller sample size ultimately leads to (1) an inability to translate findings to a broad understanding of galaxies, and (2) a dearth of statistical sub-samples of galaxies in unique positions of evolution (mass, environment, star-forming properties, etc.). iEDGE \citep{Colombo25-iEDGE} is a recent survey that strikes a balance between the resolution of the mm data and sample size. They do so by taking advantage of aperture-corrected and non-aperture-corrected CO luminosities to make an assessment of the global and central gas reservoirs (paired with resolved optical spectroscopy) for a large statistical sample (643 in this case). 

The Atacama Large (sub-)Millimeter Array (ALMA) offers the best combination of resolution and sensitivity for observing in the mm to sub-mm regime. From first light in 2011, the array has been operating for over a decade. The ALMA Science Archive has therefore amassed a wealth of resolved mm data. However, ALMA alone does not provide optical information from which key properties for understanding galaxy evolution can be derived; additional optical data must supplement the ALMA archive.

In this work, we match the ALMA archive to the SDSS, which provides optical spectroscopy and photometry, to generate a large sample of ``semi-resolved'' (here meaning the ability to probe the inner and outer regions of a galaxy separately) galaxies with matched optical and mm data. The SDSS uses central fibres for their spectroscopic measurements and provides photometric fits in multiple bands, allowing for an independent assessment of both the inner and outer stellar properties of galaxies. Combined with resolved CO emission maps from ALMA, we produce a sample of galaxies with the stellar and gaseous components individually distinguished in the inner and outer regions of the galaxies. We call the sample the SDSS-ALMA Legacy Value Archival Gas Exploration (SALVAGE) and it fills a niche in terms of sample size and resolution between that of the large but unresolved xCOLDGASS and the kpc-resolution IFU programs such as ALMaQUEST and EDGE-CALIFA. In this work, we introduce SALVAGE (Section \ref{Methods}) and use it to provide a semi-resolved perspective on how the thoroughly-tested global scaling relations (and deviations from them) manifest (Section \ref{allresults}). Furthermore, we discuss our results within the context of previous works and the prospect of using the distribution of molecular gas within galaxies to distinguish between different mechanisms altering the global star-forming status of galaxies (Section \ref{discussion}). Finally, our conclusions are summarized in Section \ref{Summary}. Throughout, we assume a flat $\Lambda$CDM cosmology with $\Omega_\text{M} = 0.3$, $\Omega_\Lambda = 0.7$, and H$_0 = 70$ km s$^{-1}$ Mpc$^{-1}$.
\section{SALVAGE Data Collation and Processing}
\label{Methods}

    To investigate the exchange between gas and stars and their role in galaxy evolution, we require both optical and mm data. We have thus assembled a large sample of semi-resolved observations. Moreover, we are able to do so using purely archival data from SDSS and ALMA. While the optical data in SDSS have been reduced and compiled into catalogues, the ALMA data are uncalibrated and unprocessed. In this section, we describe the collation of the optical data from various catalogues, the processing of mm data into science-ready products, and the combination of the two into a cohesive and publicly-available catalogue.

    \subsection{Data Products from the Sloan Digital Sky Survey}
    \label{sdssdata}

    \subsubsection{Overview}

    The SDSS is constructed using a dedicated 2.5 m telescope located at the Apache Point Observatory in New Mexico. It is equipped with a mosaic charged-coupled device (CCD) camera to obtain images in five optical bands, and two fibre-fed optical spectrographs to obtain spectra \citep{York00}. The seventh data release of the SDSS \citep[DR7;][]{SDSS-DR7} includes optical spectra and \emph{ugriz} broadband photometry for over 900,000 galaxies\footnote{\url{https://classic.sdss.org/dr7/}}. Although DR7 has been superceded by additional data releases, DR7 marks the end of the main galaxy survey and remains a standard resource for the community thanks to large ancillary catalogues (Section \ref{mpajhu}).
    
    The SDSS spectra have a spectral resolution of R $=\lambda/\Delta\lambda\thicksim$ 800-2200 over the wavelength range 3800-9200 Å and are collected from optical fibres which have an on-sky aperture of three arcseconds. Thus, the spectra only include light from the central $\thicksim$500 pc for lower-redshift ($z \thicksim 0.025$) galaxies and from the central $\thicksim$6 kpc for higher-redshift ($z \thicksim 0.25$) galaxies in the survey. The SDSS photometry has a median 5$\sigma$ point source depth of 22.15, 23.13, 22.70, 22.20, and 20.71 in the \emph{u}-, \emph{g}-, \emph{r}-, \emph{i}-, and \emph{z}-bands, respectively. The photometric and spectroscopic products used in our work are described in more detail below.

    \subsubsection{SDSS Photometry Catalogues}
    \label{sdss-phot}

    SDSS released catalogues of photometric sources\footnote{\url{https://cas.sdss.org/dr7/en/help/browser/browser.asp?n=PhotoObjAll\&t=U}}, which include magnitudes measured using different conventions as well as simple structural properties to characterize galaxy size. From the SDSS photometric catalogues, we extract the following:
    
    \begin{itemize}
        
        \item \emph{Model magnitudes} are derived by using best fit exponential or de Vaucouleurs profile. Each profile and magnitude comes with the parameters describing the best fit model (ellipticity, orientation, scale radius, likelihood of the best fit model). There is a upper limit of 4 scale radii for exponential fits and 8 scale radii for de Vaucouleurs fits. Model magnitudes are the suggested global magnitudes for SDSS galaxies \citep{Salim07} and we discuss how they are used to measure global stellar mass in Section \ref{mpajhu}.
        \smallskip

        \item \emph{Fibre magnitudes} are taken from the flux within a 3 arcsecond diameter circular aperture centered on the location of the fibre. Images are convolved to 2 arcsecond seeing before the aperture photometry is applied to measure the fibre magnitudes. In Section \ref{mpajhu}, we discuss the role of fibre magnitudes in measuring stellar mass within the SDSS fibre aperture.
        \smallskip

        \item \emph{The Petrosian half light radius} is the radius that contains 50\% of the Petrosian flux. We use the $r$-band Petrosian radius as our radial unit of size.
    
    \end{itemize}

    \subsubsection{SDSS Spectroscopic Catalogues}

    The SDSS released catalogues of spectroscopic sources\footnote{\url{https://cas.sdss.org/dr7/en/help/browser/browser.asp?n=SpecObjAll\&t=U}} containing the spectroscopic redshift and emission line fluxes, as well as higher order values such as automated classification of the spectra. In this work, we use the spectroscopic redshifts and require that the spectra are classified as galaxies or quasars.

    \subsubsection{Ancillary Data Products from MPA-JHU}
    \label{mpajhu}
    
     The Max-Planck-Institute for Astrophysics–John Hopkins University (MPA-JHU) catalogues\footnote{\url{https://wwwmpa.mpa-garching.mpg.de/SDSS/DR7/}} provide flux measurements of up to 12 emission lines, photometric stellar mass estimates, and hybrid photometry/spectral index SFR estimates \citep{K03, Tremonti04, Brinchmann04}. While the original works are based on SDSS DR4 data, the most up-to-date catalogue is for SDSS DR7 and includes some changes from the methods discussed in \citet{K03} and \citet{Brinchmann04}. Here we summarize key details regarding the data products used in this work:

    \begin{itemize}
    
        \item \emph{Emission line flux measurements} are taken directly from the public MPA-JHU raw data catalogue\footnote{\url{https://wwwmpa.mpa-garching.mpg.de/SDSS/DR7/raw\_data.html}}. Fluxes in the catalogue are corrected for foreground Galactic reddening following \citet{Odonnell94}. However, an additional correction to the emission line fluxes in the catalogue is applied, accounting for the internal galactic reddening of the target galaxy using a Milky Way extinction curve, as parameterized by \citet{Cardelli89}.
        \smallskip
        
        \item \emph{Stellar mass estimates} are determined following similar methods as \citet{K03}, in the sense that a best-fit model spectrum from \citet{BC03} is used to estimate the mass-to-light ratio (among other properties) of each galaxy. The stellar mass is estimated by multiplying the mass-to-light ratio by the luminosity derived from the photometry. The key difference is that \citet{K03} used spectral indices from the fibre spectrum, while the updated DR7 catalogue uses the optical photometry; using optical photometry allows for consistent and separate fits to the fibre and total photometry and has no systematic offset at $M_\star>10^9M_\odot$\footnote{\url{https://wwwmpa.mpa-garching.mpg.de/SDSS/DR7/mass\_comp.html}}. Multiplying the mass-to-light ratio inferred from fibre photometry by the luminosity within the fibre gives the stellar mass within the 3" fibre aperture which we refer to as \emph{inner stellar mass} ($M_{\star\text{, inner}}$). Multiplying the mass-to-light ratio inferred from ModelMag photometry by the total luminosity gives the stellar mass of the entire galaxy which we refer to as \emph{total stellar mass} ($M_{\star\text{, total}}$). The difference between the two traces the mass in the outer annulus of the galaxy which we refer to as \emph{outer stellar mass} ($M_{\star\text{, outer}}$). The stellar masses have a typical uncertainty of $\thicksim$0.1 dex.
    
        \smallskip
        
        \item \emph{Star formation rates} from the MPA/JHU catalogue are delivered for both star-formation within the fibre aperture and the total extent of the galaxy. Within the SDSS fibre, SFR is measured on the basis of many emission lines (specifically H$\alpha$, H$\beta$, [OIII]$\lambda5007$, [NII]$\lambda6584$, [OII]$\lambda3727$, and [SII]$\lambda6716$), with the greatest weight carried by H$\alpha$ \citep[as in ][]{Brinchmann04}. For cases where there is no H$\alpha$ emission or when there is AGN contamination, SFRs are derived from the 4000 Å break, calibrated using the H$\alpha$ sSFR estimates of star-forming galaxies. We refer to SFRs calculated within the fibre as \emph{inner SFR} (SFR$_\text{inner}$). The \emph{outer SFR} (SFR$_\text{outer}$) is estimated by isolating the optical photometry outside the fibre (cModelMag - fiberMag in each \emph{ugriz} band) and performing model fitting following the method of \citet[][]{Salim07}. The SFR measured outside the fibre is therefore independent from that measured within the fibre and is added to the SFR inside the fibre to get the \emph{total SFR} (SFR$_\text{total}$).

    \end{itemize}

    Since the MPA/JHU catalogues independently measure the SFR for the inner region with spectroscopy and the outer region with photometry, the total SFRs are obtained by combining SFR values that have been derived from different methods. Although \citet{Salim16} has previously shown that the SFRs measured using spectroscopic and photometric methods are consistent for star-forming galaxies, we test whether combining multiple SFR methods has an impact on the conclusions presented in this work. We compare the total SFRs used in this work to the purely photometrically derived SFRs measured in \citet{Salim16} in Appendix \ref{GSWLC}, and find that our conclusions are robust under changes to global SFR estimates.

    \subsubsection{GalaxyZoo Morphology Catalogue}

    Although our optical data come primarily from SDSS, SALVAGE has complete coverage within the Dark Energy Camera Legacy Survey \citep[DECaLS;][]{Dey19}, which provides slightly deeper (23.54 mag 5$\sigma$ point source depth in the $r$-band) and higher-resolution (1.18" FWHM) imaging than SDSS, which allows for a more reliable estimate of morphology and faint features \citep{Wilkinson24, Bickley24-mergeridentification}. We use the Galaxy Zoo DESI morphology catalogue \citep{Walmsley23} trained on DECaLS Galaxy Zoo responses to predict key morphological features. In particular, we use the following parameters:

    \begin{itemize}
    
        \item \texttt{disk-edge-on\_yes\_fraction} is the fraction of votes indicating if there is a stellar disk and that it is viewed edge-on. In this work, we use a threshold of \texttt{disk-edge-on\_yes\_fraction} $ > 0.5$ to identify edge-on galaxies.
        \smallskip

        \item \texttt{merging\_minor-disturbance\_fraction} and \texttt{merging\_major-disturbance\_fraction} are the fraction of votes indicating that a minor or major merger has disturbed the stellar morphology, respectively. 
        \smallskip

        \item \texttt{bar\_weak\_fraction} and \texttt{bar\_strong\_fraction} are the fraction of votes indicating that a weak or strong stellar bar is present, respectively.
        
    \end{itemize}
     
    In total, there are 674,633 SDSS galaxies with the data products described in this Section. 568,876 have a declination less than +47$^\circ$ (corresponding to an elevation of 20 degrees at the ALMA site) and are thus possibly visible by ALMA.

    \subsection{Data Products from the ALMA Science Archive}
    \label{ALMAData}

    \subsubsection{Overview}

    ALMA is a telescope array operating in the Chilean Atacama desert consisting of 66 antennas assembled into three different arrays: the ``main array,'' the ``7 m array,'' and the ``total power array.'' In this work, we use archival data from the main array, which accounts for 50 antennas, each with dishes 12 m in diameter, that operate together as an interferometric telescope. The antennas can be arranged in different configurations to change the resolution and maximum resolvable scale of the observation. The array is equipped with ten different receiver bands that allow users to select their wavelength range of interest (ranging from 0.32 mm to 3.6 mm) and spectral resolution. At the redshift range considered in this work, the 115 GHz $^{12}$CO~(1--0) emission line (henceforth CO) is covered by ALMA band 3 ($\thicksim$2.7 mm).

    ALMA began its first official cycle of observations in 2013, and as of writing, it is now executing its eleventh observing cycle. Each day, ALMA generates $\thicksim1$ TB of data\footnote{\url{https://www.almaobservatory.org/en/factsheet/}} that are stored in the ALMA Science Archive\footnote{\url{https://almascience.nrao.edu/aq/}}. As of October 16 2024, the entirety of the ALMA Science Archive contained 40,934 unique targets. Recent works have demonstrated the power of conducting extragalactic research with data from the ALMA Archive \citep[e.g.][]{Rizzo23, Bollo24, Bertola24, Ledger24, Ledger25}. However, no one has undertaken the process of compiling a complete census of extragalactic observations publicly available in the ALMA archive.

    In this section, we describe how we match the SDSS to the ALMA archive, then download, calibrate, image, and process the ALMA archival data. A large-scale archival project of this kind requires large amounts of disk storage for large volumes of raw and reduced data, substantial random access memory (RAM) given the size of individual cubes that must be generated and loaded into memory, multiple computer processing units (CPUs) to minimize clock time by reducing several targets at once, and a suite of software (see Section \ref{Calibration}). The Canadian Advanced Network for Astronomy Research (CANFAR\footnote{\url{https://www.canfar.net/en/}}) Science Platform is uniquely equipped to address all of these demands. As such, the work described here is done on the CANFAR Science Platform, which we will refer to throughout. 

    \subsubsection{The SDSS-ALMA Crossmatched Sample}
    \label{Sample}

    Since we seek complete coverage of SDSS and ALMA, our parent sample is comprised of SDSS DR7 (the main galaxy sample) with MPA/JHU masses and star-formation rates, that is possibly observable by ALMA (i.e. with a declination less than 47 degrees). There are 568,876 SDSS galaxies that meet this cut. 

    We query the ALMA Science Archive using \texttt{pyVO} and match to SDSS using \texttt{match\_to\_catalog\_sky} function from \texttt{astropy} \citep{Astropy1, Astropy2, Astropy3}, which yields 739 unique matches with an on-sky matching tolerance of four SDSS $r$-band Petrosian radii and where the data product type is a spectral cube meeting the quality requested by the original PIs (QA2 PASS and \texttt{scan\_intent=TARGET}). From here, we further constrain our sample as follows:

    \begin{itemize}
    
        \item Must have $^{12}$CO~(1--0) line coverage according to the redshifted 115.27 GHz line (using the SDSS spectroscopic redshift) falling within the reported bandwidth. 517 of the 739 SDSS-ALMA matches meet this criterion.
        \smallskip

        \item Must have a reported spatial resolution smaller than the SDSS central fibre (3 arcsec). This spatial resolution allows us to independently resolve the inner region, which we define to be the SDSS fibre (to match to the optical data). 324 targets from the previous step meet this criterion.
        \smallskip

        \item Must have a maximum resolvable scale larger than 1.5 Petrosian radii so as to not resolve out significant flux. 307 from the previous step meet this criterion.
        \smallskip

        \item Must not be the inferior of two duplicate observations. 9/307 targets from the previous step are duplicate observations, generally separated by many cycles (e.g. Cycle 1 observations being replaced by Cycle 9) or simultaneous observations from different projects in the same cycle. In cases where the duplicates span multiple cycles, we take the newer observation. In cases where the observations are from the same cycle, we take the observation with the better line sensitivity, and in cases where there are multiple observations of the same target in the same project, we use all the observations. 298 from the previous step meet this criterion.

    \end{itemize}
    
    After the SDSS crossmatch to the ALMA archive and sample selection, we move forward with processing a sample of 298 targets.

    \subsubsection{Calibration}
    \label{Calibration}

    Starting in Cycle 3, ALMA began to deliver calibrated data using a pipeline developed in-house with Common Astronomy Software Applications \citep[CASA;][]{CASA22}. Although the data stored on the ALMA archive are uncalibrated, a script is available that ``restores'' the pipeline calibrations.

    For observations from Cycle 3 or later, we restore the pipeline calibrations using the script included with the downloaded data. The calibration restoration script must be run using the most recent version of CASA available \emph{at the time of observation}. In total, we required 13 different versions of CASA (ranging from v4.2.2 to v6.5.4) that are pre-installed on the CANFAR Science Platform. Observations in our sample that were observed in Cycles 0, 1, or 2 were never pipeline calibrated, and thus no script is provided to either restore or run pipeline calibrations. For the 46 such targets from these first three cycles, the calibrated measurement sets were provided to us by the ALMA Help Desk staff.

    \subsubsection{Imaging}
    \label{Imaging}

    ALMA visibilities were imaged to generate the final continuum-subtracted datacubes, deconvolved from the dirty beam. Throughout the imaging process, there are a number of choices that must be made that can have a significant impact on the final results. Motivated by the need to image hundreds of observations efficiently and in a way that is reproducible by others, we automate our imaging using the pipeline\footnote{\url{https://github.com/akleroy/phangs\_imaging\_scripts?tab=readme-ov-file}} developed by the PHANGS-ALMA team\footnote{\url{https://sites.google.com/view/phangs/home}}\citep{Leroy21}. Although the details of the pipeline are described in detail in \citet{Leroy21}, we will briefly describe the aspects relevant to this work.

    The PHANGS pipeline is managed using `key' files to describe all inputs to the pipeline including target names, positions and velocities, the location of the calibrated measurement sets for each target, and configurations for the cleaning and post-processing of the cubes. The pipeline is run from a Python script that calls the individual functions of the pipeline. 

    First, the pipeline stages the data for imaging. The science data are isolated from other data in the measurement set (i.e. calibrators) using the \texttt{CASA} task \texttt{split}. However, we found that it was necessary to conduct our own \texttt{split} call before passing to the PHANGS-ALMA pipeline to isolate specific targets from measurement sets that included multiple science targets. The pipeline then conducts a continuum subtraction using the \texttt{CASA} task \texttt{uvcontsub}. The continuum is fit with an order 0 polynomial using only the spectral window that contains the CO line (and separately for each individual case, if there are multiple observations). The data are then regridded to a common velocity grid that is unique to each galaxy using \texttt{CASA} task \texttt{mstransform}. We select a target velocity resolution of 25 km/s (using \texttt{channel\_kms} in the key files) to ensure the spectral line is well sampled but also to maximize signal-to-noise. The PHANGS-ALMA pipeline does not force this exact velocity resolution, but rather identifies an integer multiple of the native spectral resolution that best matches the user-desired target velocity resolution, without exceeding it. We choose to include 1000 km/s of spectral coverage on either side of the redshifted CO line frequency to ensure there are sufficient channels to compute robust noise statistics, but also minimize data volume.
    
    The pipeline then images the data with the \texttt{CASA} task \texttt{tclean} in a two-step process. In the first imaging step, the pipeline uses multiscale cleaning with the goal of producing a reliable map to use as input to the second stage of cleaning. Since the typical beam size in SALVAGE ($\thicksim 1-3$") is similar to the PHANGS 12 m data ($\thicksim 1-2$"), we use the same cleaning scales of 0, 1, 2.5, 5.0 arcseconds (but found that our results did not change significantly with deviations from this). In the second imaging step, the pipeline uses the standard single-scale CLEAN algorithm \citep{Hogbom74} to produce a final deconvolved cube. In this instance, the cube is cleaned down to a signal-to-noise of 1 and the PHANGS pipeline forces frequent major cycles using multiple \texttt{tclean} calls that iteratively add the number of allowed clean components by a factor of two until the model flux changes by less than 1\% between \texttt{tclean} calls. Throughout the cleaning process, visibilities are weighted using a Briggs scheme with robust = 0.5, the cell size is chosen to be a round number that samples the minor axis of the beam by $\gtrsim4$, and the primary beam cutoff is set to 0.25. For any other details, we again refer the reader to \citet{Leroy21}.
    
    Our implementation\footnote{\url{https://github.com/sj-wilkinson/SALVAGE/}} of the PHANGS pipeline is designed to run on the CANFAR Science Platform. It downloads the data with \texttt{ALMiner} \citep{ALMiner}, calibrates the data, prepares the key files for the PHANGS pipeline, and runs the PHANGS pipeline, for each galaxy.

    \subsubsection{Computing molecular gas mass}
    \label{Post-processing}

    The cleaned CO spectral cubes allow us to measure the molecular gas mass on the same scales as the SDSS optical data products (Section \ref{sdssdata}). To this end, we extract ``inner,'' ``outer,'' and ``total'' spectra from the cleaned cube by spatially summing the spectra from pixels of the cube that fall within apertures equal to the SDSS data: for the inner spectrum, we include pixels within the central 3"-diameter aperture; for the outer spectrum, we include pixels that fall within an annulus with a radial extent from 1.5" to the 4 scale lengths of the best fit SDSS photometric model; for the total spectrum, we include all pixels within a radius equal to 4 scale lengths of the best fit SDSS photometric model. 

    Next, we measure the CO line emission ($S_\text{CO}$) from each spectrum using \texttt{spectral-cube}. To maximize signal-to-noise in the flux measurement, we seek to sum over only the channels that contain flux from the source. We do this using the ``full width at zero intensity" method (FWZI) also employed in VERTICO \citep{Brown21}. Starting from the peak flux channel of the spectrum, this method includes all contiguous channels with flux above zero (allowing for the inclusion of asymmetric velocity structure). In the case of a continuum-subtracted spectrum with well-behaved noise properties, a centrally-peaked spectrum (as is the case for most ``inner'' spectra) crosses zero within 1-2 channels of when the spectrum becomes dominated by noise. For the ``outer'' spectra, the FWZI process is repeated thrice, starting from the three highest flux channels. After three iterations, all channels identified as containing flux are used to measure the outer CO line emission. This works well to include cases where strong flux may be separated significantly in velocity space (due to spatially summing over both sides of a rotating disk), without strong contamination from noise (in some rare cases, the third peak may occur in a noise spike, but only includes one channel because the noise promptly falls below zero in neighbouring channels).

    The spectrum root mean square (RMS) noise is computed from the standard deviation measured from the outer 200 km/s on either side of the summed spectrum; therefore, the noise is computed over the $\thicksim$16 channels furthest from the emission line. However, these channels represent the noise properties of the summed spectrum at the same spatial location(s) across the cube from which the spectra were extracted. In the case of non-detections of $S_\text{CO}$ where the signal-to-noise ratio is less than 5, we compute an upper limit assuming a line width of 300 km/s \citep[following xCOLD GASS;][]{Saintonge17} as $\sigma\Delta V\sqrt{N}$, where $\sigma$ is the RMS noise in Jy, $\Delta V$ is the channel width in km/s and $N$ is the number of channels equal to 300 km/s (i.e. about 12 channels).

    We convert the observed line fluxes ($S_\text{CO}$) to CO integrated line luminosities ($L'_\text{CO}$), following equation (3) from \citet{Solomon05}:

    \begin{equation}
        L'_\text{CO} = 3.25\times10^7S_\text{CO}\nu^{-2}D_L^2(1+z)^{-3},
        \label{LCOeq}
    \end{equation}

    \noindent where $\nu$ is the expected frequency of the line according to the SDSS spectroscopic redshift in GHz and $D_L$ is the luminosity distance in Mpc.

    The CO integrated line luminosity is converted to a molecular gas mass estimate, $M_\text{mol}$, using a constant conversion factor:

    \begin{equation}
        M_\text{mol} = \alpha_\text{CO}L'_\text{CO},
        \label{MH2eq}
    \end{equation}

    \noindent where $\alpha_\text{CO} = 4.35$ M$_\odot$ (K km s$^{-1}$ pc$^{2}$)$^{-1}$ \citep{Schinnerer24}. In Section \ref{discussion_alpha}, we test how our results would change with a different $\alpha_\text{CO}$ prescription and find that our qualitative results hold. 

    Our end result is molecular gas masses extracted from the inner, outer, and total regions that match the spatial scales of the SDSS optical data products, which we call $M_\text{mol, inner}$, $M_\text{mol, outer}$, and $M_\text{mol, total}$. These data and their optical counterparts are provided in the machine-readable Table \ref{SubmmTable}.
    
    \begin{table*}
    \centering
    \begin{tabular}{|r|c|c|c|c|c|c|c|c|c|c|c|}

SDSS DR7 Object ID & M$_{\star\text{, total}}$ & M$_{\star\text{, inner}}$ & M$_{\star\text{, outer}}$ & SFR$_\text{total}$ & SFR$_\text{inner}$ & SFR$_\text{outer}$ & M$_{\text{mol, total}}$ & M$_{\text{mol, inner}}$  & M$_{\text{mol, outer}}$ \\

& log$\left(\frac{\text{M}_\star}{\text{M}_\odot}\right)$  & log$\left(\frac{\text{M}_\star}{\text{M}_\odot}\right)$  & log$\left(\frac{\text{M}_\star}{\text{M}_\odot}\right)$  & log$\left(\frac{\text{SFR}}{\text{M}_\odot \text{/yr}}\right)$ & log$\left(\frac{\text{SFR}}{\text{M}_\odot \text{/yr}}\right)$ & log$\left(\frac{\text{SFR}}{\text{M}_\odot \text{/yr}}\right)$ & log$\left(\frac{\text{M}_{\text{mol}}}{\text{M}_\odot}\right)$ & log$\left(\frac{\text{M}_{\text{mol}}}{\text{M}_\odot}\right)$ & log$\left(\frac{\text{M}_{\text{mol}}}{\text{M}_\odot}\right)$\\ 

\hline





587741490891325518 & 11.0 & 10.2 & 10.9 & 0.9 & 0.7 & 0.3 & 10.2 & 9.5 & 10.1 \\
587722984441250023 & 10.2 & 9.9 & 9.9 & 0.2 & 0.1 & -0.4 & 9.3 & 9.2 & 8.7 \\
588015508213989555 & 10.5 & 9.6 & 10.4 & 0.5 & -0.2 & 0.4 & 9.7 & 8.8 & 9.7 \\
587741421099155681 & 9.8 & 9.0 & 9.7 & 0.1 & -1.0 & 0.0 & 9.0 & 8.1 & 9.0 \\
587727221396013696 & 10.3 & 9.5 & 10.2 & 0.3 & 0.1 & -0.1 & 9.4 & 8.7 & 9.3 \\
587742903401119752 & 9.5 & 9.1 & 9.3 & -0.7 & -1.2 & -0.8 & 7.6 & 7.0 & 7.5 \\
587731512619696292 & 10.3 & 9.8 & 10.1 & 0.2 & 0.2 & -0.5 & 9.4 & 8.6 & 9.3 \\
587731173842027508 & 11.1 & 10.5 & 11.0 & 1.1 & 0.5 & 0.9 & 10.0 & 9.4 & 9.9 \\
587722984438235345 & 10.6 & 10.0 & 10.4 & 0.4 & -0.3 & 0.3 & 9.8 & 8.9 & 9.7 \\
587731511542022280 & 10.5 & 9.7 & 10.4 & 0.1 & -0.5 & -0.0 & 9.1 & 8.3 & 9.0 \\

 \vdots & \vdots & \vdots & \vdots  & \vdots  & \vdots  & \vdots  & \vdots  & \vdots  & \vdots

    \end{tabular}
    \caption{Key semi-resolved products for the first ten galaxies in SALVAGE. The full table includes observational quantities from the ALMA data reduction (e.g. $S_\text{CO}$, $L'_\text{CO}$), multiwavelength parameters (e.g. SFE, $f_\text{gas}$) described in Section \ref{mwdata}, and associated errors for these quantities for all 277 galaxies in SALVAGE.}
    \label{SubmmTable}
\end{table*}

    \begin{figure*}
        \centering
        \includegraphics[width=1\linewidth]{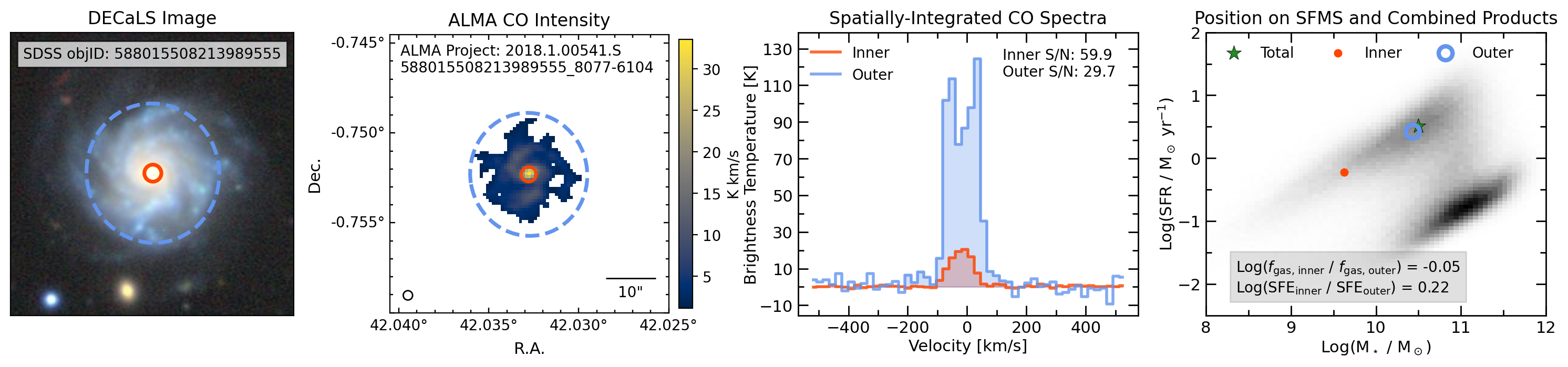}
        \caption{Example data products for one target (SDSS object ID: 588015508213989555) in SALVAGE. \emph{Left panel:} The DECaLS \emph{gri} colour image with the SDSS fibre aperture overlaid in red and the approximate extent of the SDSS photometric model in blue. \emph{Centre-left panel:} the ALMA CO moment 0 map with the colour corresponding to the CO line strength according to the colorbar to the right. The size of the circularized beam and a scale bar representing 10" are overlaid in black along the bottom. The ALMA project code and target name (as per that project) are along the top. \emph{Centre-right panel:} the spatially-integrated inner (red) and outer (blue) spectra extracted from the cleaned ALMA cube within the SDSS fibre aperture (red circle in centre-left panel) and the annulus between the fibre and photometric limit (blue dashed circle in centre-left panel), respectively. Channels for which the area under the line is shaded represent those identified as contributing flux. The signal-to-noise ratio of the lines is shown in the top right. \emph{Right panel:} the inner (red), outer (blue), and total (green) locations on the SFR-M$_\star$ plane with all SDSS galaxies in shaded black for reference. To demonstrate the combination of semi-resolved optical and mm data, the ratio between the inner and outer $f_\text{gas}$ (see Eq. \ref{fgas}) and SFE (see Eq. \ref{SFE}) are written in text along the bottom of the panel.}
        \label{ReductionExample}
    \end{figure*}

    In Figure \ref{ReductionExample}, we present an example galaxy with visualizations of its constituent data products from SDSS and ALMA. In the left panel of Figure \ref{ReductionExample}, we show the 50" $\times$ 50" \emph{gri} colour image drawn from DECaLS \citep{Dey19}. Overlaid in red is the 3" SDSS fibre aperture located at the centre of the galaxy and overlaid in blue is the 4 scale length extent of the SDSS photometric model fit which serves as an approximate maximum extent of the SDSS model photometry. In the centre left panel, we show the moment 0 map of the cleaned datacube (with a circularized beam for visualization purposes only) which is output from the PHANGS-ALMA pipeline. In the centre-right panel, we show the spatially-integrated inner (red) and outer (blue) CO spectra, extracted from the region within 3" SDSS fibre aperture and between the fibre and maximum extent of the photometry, respectively. The channels with shaded colour under the spectrum are those included in the flux measurement of the inner and outer regions. In the right panel, we show how the inner (red point), outer (blue annulus), and total (green star) can be placed on the SDSS SFMS and demonstrate additional combined SDSS and ALMA semi-resolved data along the bottom (described in Section \ref{mwdata}).

    \subsubsection{Quality Assurance}
    \label{qa}

    The optical image from DECaLS, the imaged CO spectral cube from ALMA, and the inner/outer/total integrated spectra were visually inspected to ensure that the data are suitable for our work. During this visual inspection, we found two common pitfalls. First, in five cases, the SDSS fibre was placed off-centre, away from the inner region of the galaxy. Second, we found nine cases where the CO spectra exhibit an unreliable noise structure. There are also seven galaxies for which either the calibration or imaging failed and could not be recovered. In total, we have 21 cases for which the data could not be included in SALVAGE. This quality assurance check brings our sample size to 277.

    19 targets in SALVAGE have previous unresolved global CO measurements from xCOLD GASS (14 targets in common) and MASCOT (5 targets in common). We used this overlap to perform a consistency check between the CO luminosities in the literature and those computed from observations on the ALMA archive and reduced using the PHANGS-ALMA pipeline in this work. In 12/14 of the overlapping targets, we measure a CO detection with S/N$>5$. The median difference between our measurement of $L'_\text{CO}$ and those in xCOLD GASS/MASCOT for these 12 detections is 0.02 dex ($\thicksim$5\%) with a scatter around zero of $\thicksim$0.09 dex, which is similar to the typical error on xCOLD GASS luminosities ($\thicksim$0.07 dex). The two cases for which we do not recover a CO detection in SALVAGE are not outliers in terms of resolution or maximum resolvable scale -- the ALMA observations are just not as deep. Since our luminosities (and thus our molecular gas mass estimates) are of similar magnitude, with no obvious systematic offsets, we conclude that our methods are consistent and that we are not resolving out significant flux at the selected resolutions.

    \subsubsection{Description of the SALVAGE Sample}

    SALVAGE is an opportunistic and heterogeneous ensemble of 277 galaxies built from the individual proposals stored in the ALMA archive. The majority of the SALVAGE sample ($\thicksim$ 55\%) is comprised of three programs: project code 2021.1.01089.S (PI: Cortese), which primarily targets main sequence galaxies; ALMaQUEST (PIs: Lin, Ellison, Pan), which targets starbursting, main sequence, and green valley galaxies at the resolution of the MaNGA IFU \citep{Lin20, Ellison24-aqpde}; and CO-CAVITY (PI: Espada), which targets galaxies in low-density environments \citep{Rodriguez24}. Based on their stated goals, these projects should supply a substantial number of representative galaxies, without strong selection biases. 
    
    There are other significant contributing projects: 2013.1.00530.S (PI: Ibar), which targets star-forming and dusty galaxies on the basis of far-IR emission \citep[see VALES, e.g.][]{Villanueva17}; 2017.1.00601.S (PI: Saintonge), which targets main sequence galaxies with ionised gas outflows \citep{Hogarth21, Hogarth23}; and 2013.1.00115.S (PI: Espada), which targets low-mass star-forming galaxies. There are 30 other individual ALMA projects included, each contributing between 1 and 11 targets to the SALVAGE sample. A full list of contributing ALMA project codes can be found in the Acknowledgments.

    \begin{figure*}
        \centering
        \includegraphics[width=1\linewidth]{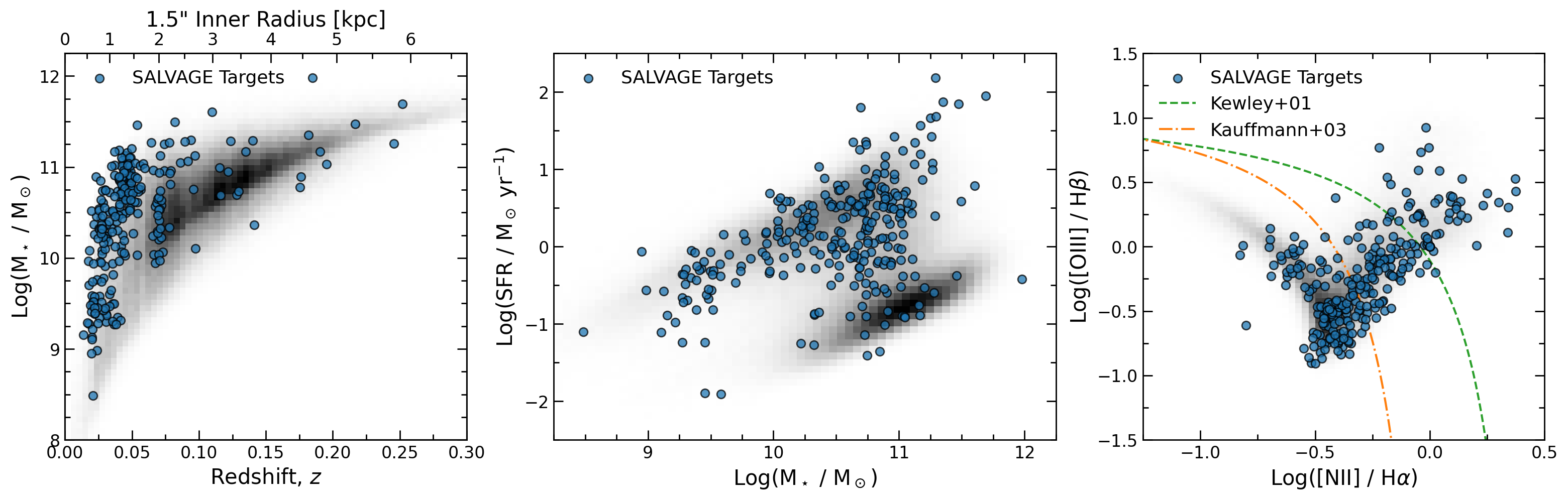}
        \caption{\emph{Left:} The stellar mass and redshift of SALVAGE targets (blue points) compared to all of SDSS (gray shading). SALVAGE is biased towards high stellar mass and low-redshift targets. \emph{Centre:} SFR and stellar mass of SALVAGE targets (blue points) compared to all of SDSS (gray shading). SALVAGE consists of a diverse sampling of galaxies above, on, and below the SFMS across a broad range of stellar masses. \emph{Right:} BPT diagram of SALVAGE targets (blue points) compared with all of SDSS. The orange dot-dash line represents the \citet{K03} AGN/SF demarcation and the dashed green line represent the \citet{Kewley01} AGN criteria. SALVAGE includes galaxies with a variety of emission line characteristics including pure star-forming galaxies, AGN, and composites.}
        \label{sample}
    \end{figure*}

     Although the galaxies garnered from the ALMA archive are heterogeneously selected and observed, the complete sample of archival galaxies spans a variety of evolution and ionization states. However, as an archival sample, it is important to be aware of the biases introduced by including many individual proposals with varying science objectives. In Figure \ref{sample}, we present several key optical properties of the archival galaxies. In the left panel, we present the stellar mass and redshift of SALVAGE targets (blue points) contrasted with the SDSS parent sample (gray shaded histogram). Objects in SALVAGE have a range of redshifts from $\thicksim0.02$ to $\thicksim0.25$ with a median of $0.044$. The redshift of each target in SALVAGE determines the physical size of the inner region probed by the 3" SDSS fibre, which is shown along the top axis of the left panel. The radii of the central region probed by the SDSS fibre size ranges from $\thicksim$0.5-6 kpc, with a median of 1.3 kpc. The left panel of Figure \ref{sample} also shows that SALVAGE is biased toward high stellar mass and low-redshift galaxies in SDSS, as these are commonly selected to be more likely to achieve detections. However, SALVAGE spans a range of masses from 10$^{8.5}M_\odot$ to 10$^{12}M_\odot$ and to redshifts as high as 0.3. The central panel of Figure \ref{sample} demonstrates that the galaxies in SALVAGE occupy a diversity of positions in the SFR-M$_\star$ plane, including galaxies above the main sequence (at high and low masses), main-sequence galaxies, green valley galaxies, and even some on the passive sequence (though significantly underrepresented). The right panel of Figure \ref{sample} shows the BPT diagram, used to assess the ionization properties of the optical spectra \citep{Baldwin81}. SALVAGE galaxies are well populated throughout the star-forming, composite, and AGN regions of the diagram.

    In summary, SALVAGE is a diverse ensemble of 277 galaxies with independent measurements of the inner and outer regions, which fills a niche in terms of sample size and resolution between that of the large unresolved surveys such as xCOLDGASS and the smaller kpc-scale resolution surveys such as ALMaQUEST.

    \subsection{Multi-wavelength Data Products}
    \label{mwdata}

    We combine the SDSS and ALMA data into a few key parameters commonly used in both global and resolved multi-wavelength studies, described here for completeness.
    
    The specific star formation rate (sSFR) describes the SFR of a galaxy, normalized by the total stellar mass (M$_\star$), computed as:

    \begin{equation}
        \text{sSFR} = \frac{\text{SFR}}{M_\star}.
    \end{equation}

    The sSFR of a galaxy is related to its position relative to the star-forming main sequence (SFMS). However, since the sSFR of the SFMS changes with mass and redshift, sSFR is an imperfect measure of the position of a galaxy relative to the SFMS. We directly measure the SFR deviation from the SFMS at fixed stellar mass and redshift, $\Delta$SFR. $\Delta$SFR is computed as the log-difference between the SFR of the target and the SFR of galaxies on the SFMS at the same stellar mass, redshift, and environment:

    \begin{equation}
        \label{dsfr}
        \Delta \text{SFR} = \log{\text{SFR}} - \log{\text{SFR}_\text{MS}},
    \end{equation}

    \noindent where $\log{\text{SFR}_\text{MS}}$ is median SFR of star-forming galaxies (as classified by the \citet{K03} classification, with S/N$>3$) within a tolerance of 0.005 in $z$, 0.1 dex in $M_\star$, and 0.1 dex in $\Sigma_5$. The local density ($\Sigma_5$) is defined as $\Sigma_5 = \frac{5}{\pi d_5^2}$, where $d_5$ is the projected distance in Mpc to the 5th nearest neighbour within 1000 km/s. We require a star-forming comparison sample of 5 or more galaxies to constrain SFR$_\text{MS}$. Therefore, in cases where at least 5 galaxies to not meet these tolerances, they are increased by 0.005, 0.1, and 0.1 until 5 galaxies are successfully matched. 

    The gas fraction ($f_\text{gas}$) is computed as the molecular gas mass inferred from the CO luminosity ($M_\text{mol}$) divided by the stellar mass:

    \begin{equation}
        \label{fgas}
        f_\text{gas} = \frac{M_\text{mol}}{M_\star}.
    \end{equation}

    The gas fraction informs about the relative quantity of gas and a galaxy's position relative to the molecular gas main sequence (MGMS). At low-$z$, typical star-forming galaxies have global gas fractions from 0.02 to 0.25 \citep{Saintonge17} . 

    Lastly, we use star-formation efficiency (SFE) to quantify the vigor with which molecular gas is being converted into stars. SFE also quantifies the relative deviation from the Kennicutt-Schmidt relation (KS). SFE is the inverse of depletion time ($\tau_\text{dep}$), which is a measure of how long star-formation could continue given the current reservoir of gas and star-formation rate (and assuming there is no replenishment of the reservoir). SFE is computed as:

    \begin{equation}
        \label{SFE}
        \text{SFE} = \frac{\text{SFR}}{M_\text{mol}} = \tau_\text{dep}^{-1}.
    \end{equation}

    SALVAGE provides SFR, $M_\star$ and $M_\text{mol}$ for the total galaxy, as well as the inner and outer regions. We can therefore compute each of the values described in this section for the inner, outer and global regions, which we denote with subscript text (e.g. SFE$_\text{inner}$).
\section{Results}
\label{allresults}

We seek a broad characterization of the exchange between molecular gas and star formation in galaxies on the star-forming main sequence and those departing from it. To this end, we specifically investigate global scaling relations for a subset of galaxies with strong CO detections ($M_\text{mol, total}$ S/N $>5$). We revisit these well-established scaling relations with a semi-resolved perspective granted by the SDSS fibre and photometry, paired with the resolved ALMA CO maps. Therefore for this work, we additionally require either $M_\text{mol, inner}$ S/N $>5$ or $M_\text{mol, outer}$ S/N $>5$ so that the ratio between the inner and outer values is at least constrained to an upper or lower limit. In principle, detections in both would be preferred, but in practice, such a selection bias removes interesting cases in which all of the CO is centrally concentrated or centrally depleted. We also remove edge-on galaxies using the DESI GalaxyZoo classification (\texttt{disk-edge-on\_yes\_fraction} $ > 0.5$) to ensure that the SDSS fibre probes the central region of the galaxy and not the inclined disk. There are 175 galaxies in SALVAGE for which a total $M_\text{mol}$ and either an inner or outer $M_\text{mol}$ are securely detected. 

\begin{figure*}
        \centering
        \includegraphics[width=1\linewidth]{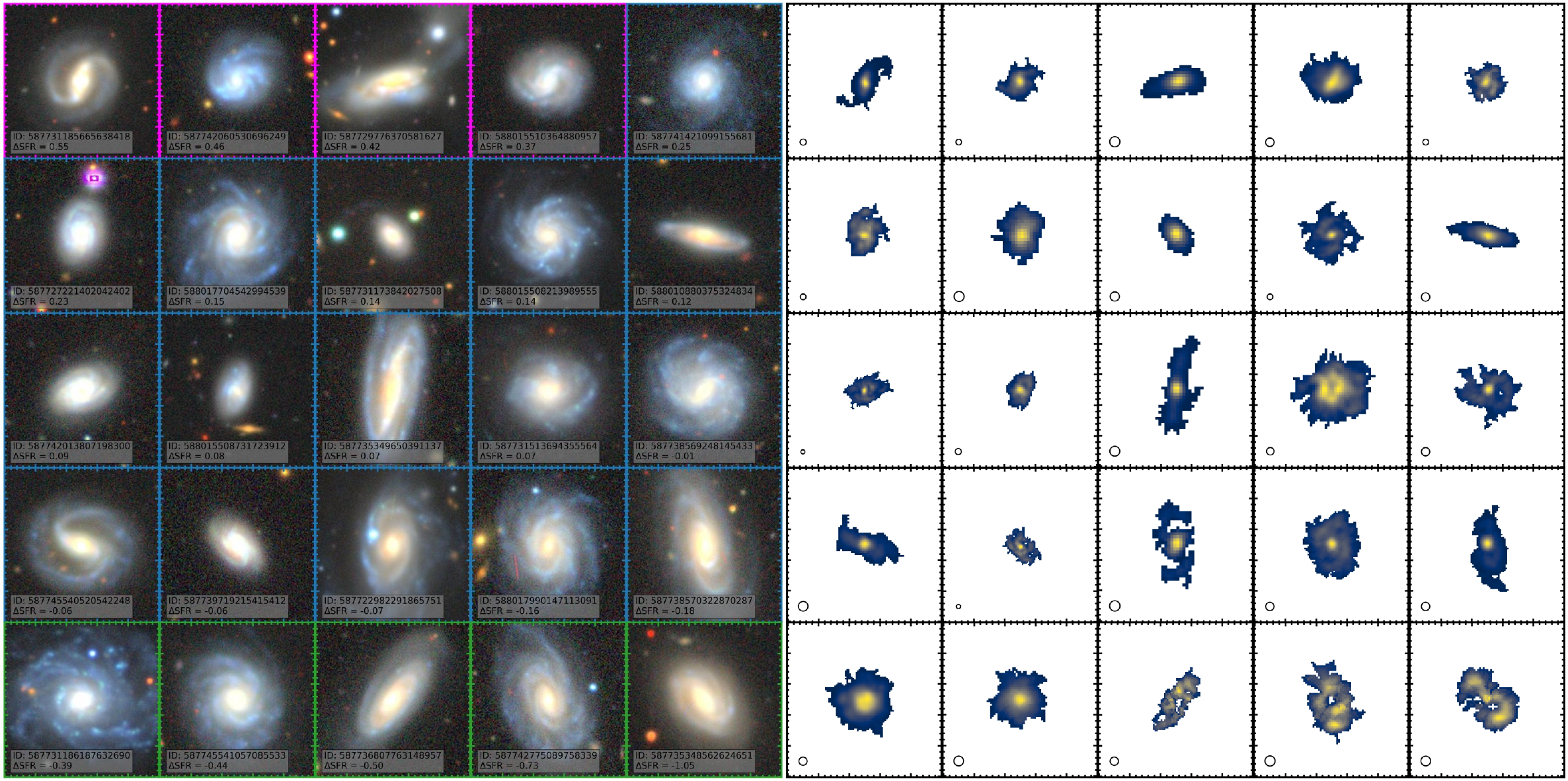}
        \caption{Images for 25 example galaxies with strong CO detections. On the left are the 50" $\times$ 50" cutouts from DECaLS (centred on the SDSS fibre) and on the right are the CO moment 0 maps (centred on the ALMA observation RA and Dec. and with a scale matched to the 50" optical cutouts). The galaxies are ordered according to their $\Delta$SFR, which is listed in the bottom left corner of the DECaLS cutouts. The circularized ALMA beam is shown as a circle in the bottom left corner of the CO moment 0 maps. However, in practice, the molecular gas mass measurements are computed from the cube with the original synthesized beam, without circularization.}
        \label{ALLexamples}
    \end{figure*}

To contextualize the sample, we present 25 examples of SALVAGE galaxies with strong CO detections in Figure \ref{ALLexamples}. On the left, we present 50"$\times$50" \emph{gri} DECaLS optical images, and on the right the CO line intensity maps, each centred on the coordinates of the SDSS central fibre. The galaxies are ordered according to their global $\Delta$SFR.

Starburst galaxies ($\Delta$SFR$>0.3$ dex) are denoted by magenta borders around their DECaLS cutouts. Their optical images tend to show either strong bars or asymmetric features indicative of a recent merger or interaction. Interestingly, the same features are mirrored by their CO profiles, although the asymmetric nature of the mergers becomes less clear. Moreover, starburst galaxies tend to have centrally concentrated CO.

Galaxies on the SFMS ($-0.3 \text{ dex} < \Delta$SFR $ < 0.3$ dex) are outlined in blue in Figure \ref{ALLexamples}. The DECaLS images reveal nothing unexpected; most star-forming galaxies are late-type spirals with bulges and bars of various sizes. The CO maps of the star-forming galaxies show the same features. However, we note that prominent central CO flux (about the size of the beam) is a nearly ubiquitous feature of star-forming galaxies (present in all but one examples in Figure \ref{ALLexamples}). 

The DECaLS images of green valley galaxies ($-1.3 \text{ dex} < \Delta$SFR$ < -0.3$ dex) are outlined in green in Figure \ref{ALLexamples}. Notably, their stellar morphologies are quite similar to that of the star-forming galaxies. However, the CO maps tell a different story: while the star-forming galaxies tend to have orderly molecular gas profiles with high central CO flux, the green valley galaxies demonstrate a diverse range of CO profiles often exhibiting azimuthal asymmetries, central holes, and otherwise disrupted molecular gas. The consistency in the optical morphologies juxtaposed by stark differences in the molecular gas structure between star-forming and green valley galaxies is an indication that the suppressed star formation in the green valley is a consequence of status of the molecular gas reservoir and its distribution throughout the galaxy.

\subsection{A semi-resolved perspective on global scaling relations}

\begin{figure}
        \centering
        \includegraphics[width=1\linewidth]{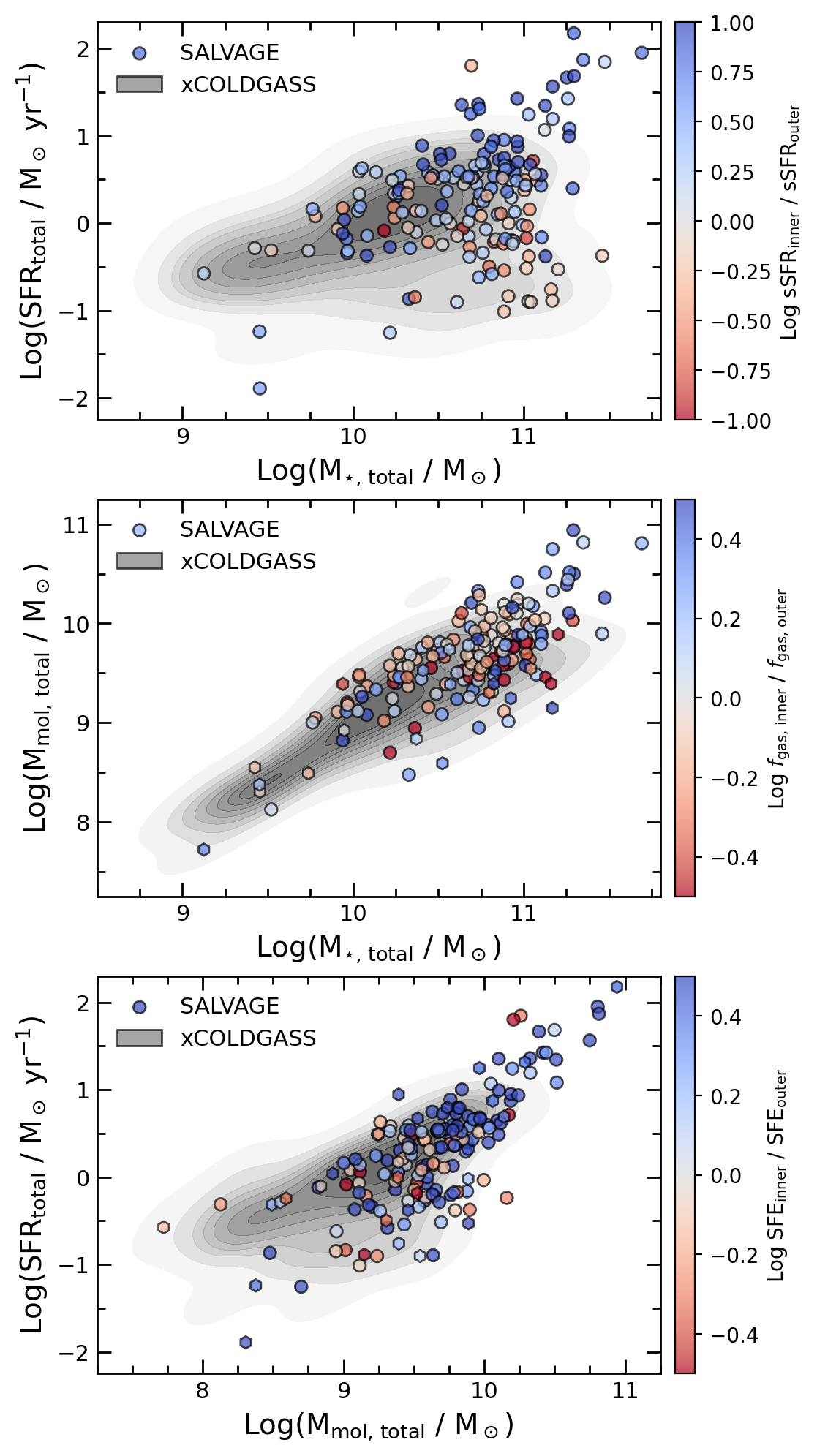}
        \caption{Key global scaling relations for the CO-detected targets in SALVAGE (coloured points) compared with detections from xCOLDGASS (gray contours) demonstrating their positions relative to the SFMS (top panel), MGMS (centre panel), and the KS relation (bottom panel). SALVAGE targets are coloured by the log ratio of the inner to the outer region of their sSFR (top panel), $f_\text{gas}$ (centre panel), and SFE (bottom panel). All points must have a 5$\sigma$ global CO detection as well as an individual inner or outer detection (also at 5$\sigma$) such that the ratio in the colour is constrained by at least an upper or lower limit; hexagon symbols are used to distinguish cases where the colour grading represents an upper or lower limit. The semi-resolved perspective provided by SALVAGE allows us to unpack these scaling relations in more detail.}
        \label{scalinglaws} 
    \end{figure}

Star-forming galaxies have been shown to form tight relations between their SFR, stellar mass, and molecular gas mass. In this section, we revisit three well-established scaling relations (namely, the SFMS, MGMS, and KS) of our CO-detected sample, with the additional context of the inner and outer semi-resolved regions to investigate why galaxies may deviate from their expected relations. Since we are investigating \emph{global} scaling relations of CO-detected galaxies, we compare to CO-detected galaxies from the large and representative (but unresolved) xCOLDGASS \citep{Saintonge17}. The SFR and stellar masses for xCOLDGASS are drawn from the same SDSS catalogue as used for SALVAGE and we compute molecular gas masses for xCOLDGASS targets using their aperture-corrected CO luminosity and a constant conversion factor of 4.35 M$_\odot$ (K km s$^{-1}$ pc$^{2}$)$^{-1}$. In Figure \ref{scalinglaws}, we present the SFMS, MGMS and KS global relations for the SALVAGE (coloured points) and xCOLDGASS (gray contours) samples. SALVAGE galaxies are coloured by the ratio of the inner and outer regions in sSFR, $f_\text{gas}$, and $\tau_\text{dep}$. In cases where the inner or outer region is not detected, an upper or lower limit is used instead.

The star-forming main sequence provides context to the global star-forming status of a galaxy. Plotting SFR$_\text{total}$ against M$_{\star\text{, total}}$ in the top panel of Figure \ref{scalinglaws}, we find that our CO-detected sample broadly falls within the scatter of xCOLDGASS. However, SALVAGE is skewed to higher stellar mass, largely because these are easier to detect in CO and are more commonly proposed to be observed with ALMA. SALVAGE has a higher propensity for starburst and green valley galaxies, probably because PIs are more likely to propose for extreme populations rather than representative samples. The ratio of sSFR in the inner and outer regions (i.e. the colour code used for the SALVAGE data points) demonstrates a large dynamic range for relative inner and outer sSFRs within galaxies. Even ``normal'' galaxies on the SFMS exhibit high variance of sSFR profiles: some have lower inner sSFRs by a factor of $\thicksim$10 (red points), others have higher inner sSFRs by a factor of $\thicksim$10 (blue points), each relative to their outer regions. Above and below the main sequence, galaxies exhibit clearer trends. Starburst galaxies frequently have much higher inner sSFR than outer sSFR. In contrast, we also find that most -- but not all -- of the galaxies in the green valley (i.e. with suppressed global SFR such that they are below the SFMS) are there primarily because of suppressed central sSFR (relative to their outer regions), in agreement with many resolved studies of sSFR profiles \citep{Ellison18, Belfiore18, Villanueva24}.

While informative of the global star-forming status of a galaxy, the SFMS is thought to be a manifestation of other, more fundamental scaling relations \citep{Lin19, Baker23}, namely, the MGMS and KS relation. Inspecting a galaxy's position relative to the global MGMS can provide context to the availability of molecular gas, relative to that which is expected given its stellar mass. In the centre panel of Figure \ref{scalinglaws}, we present the global MGMS. As expected, we find a tight relationship between the M$_\text{mol, total}$ and the M$_{\star\text{, total}}$, with only a few outliers falling below. We also note that, by comparing the location of this sequence to that of xCOLDGASS (gray contours), we find good agreement between the two, indicating reliable global molecular gas measurements.

Our data demonstrate the resilience of the global MGMS under radical changes to the molecular gas distribution within the galaxy. Within the scatter of the MGMS, galaxies commonly have equal inner and outer gas fractions (i.e. Log$(f_\text{gas, inner} / f_\text{gas, outer}) = 0$). However, there are both galaxies with inner gas fractions up to 0.5 dex above and below their outer regions, yet still fall within the scatter of the main sequence. These cases demonstrate that galaxies can remain on the global MGMS, despite having a majority of their gas in the inner or outer region (while the other is essentially devoid of molecular gas). 

Outliers from the MGMS -- both above and below -- tend to have higher gas fractions in their inner regions relative to their outer regions. Above the MGMS, this may be due to gas accretion and destabilization (e.g. from a major merger) leading to an overall increase in gas mass, but preferentially in the central regions. Below the MGMS, galaxies may have had AGN fueling and global removal or stripping preferentially from the outer regions (see Section \ref{discussion} for a more detailed discussion). The ALMA observations for galaxies with enhanced central gas fractions (relative to their outer region) but below the MGMS have maximum resolvable scales greater than five Petrosian radii; the missing molecular gas from the outer regions of these galaxies is therefore a real signal.

Rounding out our semi-resolved perspective of global scaling relations, we investigate the KS law, plotting SFR$_\text{total}$ against the M$_\text{mol, total}$ in the bottom panel of Figure \ref{scalinglaws}. Deviations from the KS relation inform us about the efficiency with which galaxies are converting their molecular gas into stars. SALVAGE targets form a tight relation, in broad agreement with observations from xCOLDGASS (e.g. similar slope). However, we note that the frequency of outliers above and below the KS relation is higher in the heterogeneously proposed SALVAGE targets, relative to the representative xCOLDGASS sample.

Most galaxies in SALVAGE have a high SFE$_\text{inner}$, relative to their outer regions. However, outliers below the KS (mostly green valley galaxies) show a variety of log$\text{SFE}_\text{inner} / \text{SFE}_\text{outer}$. Paired with the fact that these galaxies have $\thicksim0.5$dex lower SFR at fixed global molecular gas indicates green valley galaxies with globally inefficient star-formation can be induced by either central or outer inefficiency.

There are two main takeaways from investigating the semi-resolved properties of the global scaling relations. First, the structure of the SFMS is sensitive to where star formation is occurring within the galaxy. Second, the global MGMS and KS are relatively agnostic to the location of the gas and the efficiency with which it is converted into stars, respectively. Considering that the SFMS has been shown to be sensitive to the semi-resolved sSFR and that a galaxy's position relative to the SFMS is a reasonable proxy for its evolution, we now shift our attention to how the other semi-resolved properties vary across the SFMS.

    \subsection{Semi-resolved properties across the main sequence}

    In the previous subsection, we investigated the global scaling relations as a function of the underlying semi-resolved perspective (of the same relation) and how that manifests in the global relations. Our intention here is to leverage the semi-resolved approach of the scaling relations to understand the evolutionary sequence from star-forming to quiescence. Therefore, we now investigate the inner and outer properties as a continuous function of global $\Delta$SFR. While we seek continuous trends with $\Delta$SFR, we place a particular emphasis on comparing galaxies on the SFMS (i.e. $-0.3 \text{ dex}<\Delta$SFR$<0.3$ dex) to green valley galaxies (i.e. $-1.3 \text{ dex} <\Delta$SFR$<-0.3$ dex) as a test case to understand why galaxies may be departing from the SFMS.
    
\begin{figure*}
        \centering
        \includegraphics[width=1\linewidth]{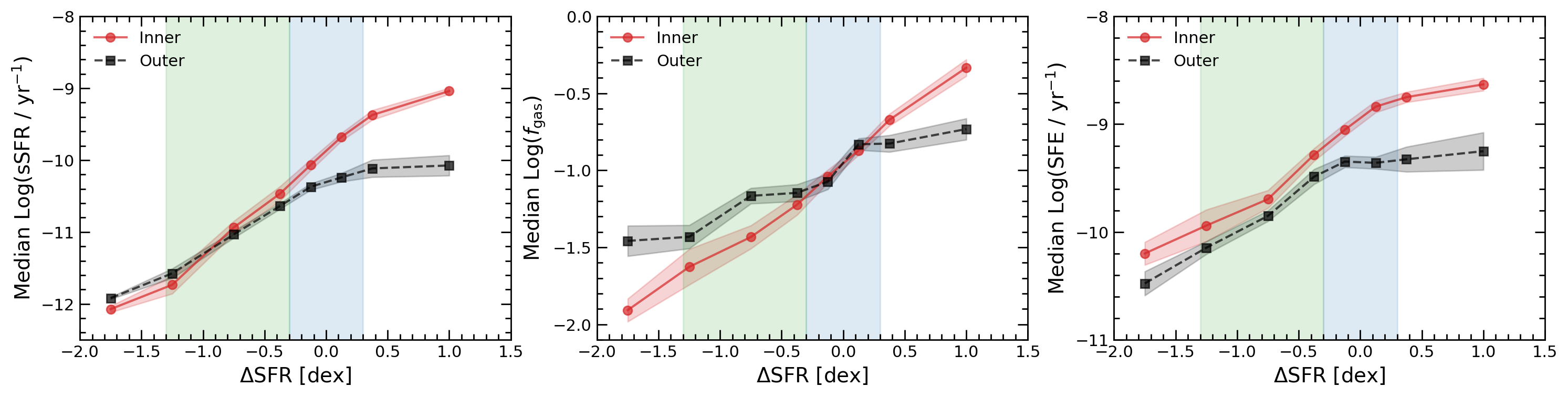}
        \caption{The median sSFR (left panel), $f_\text{gas}$ (centre panel), and SFE (right panel) of the inner (red circles) and outer (black squares) regions of SALVAGE targets  as a function of $\Delta$SFR. Each $\Delta$SFR bin spans 0.5 dex except for the region $-0.5 < \Delta$SFR$< 0.5$ where there are enough targets to reduce bins to 0.25 dex. The shaded area around the lines represent the standard error in the median ($\sigma / \sqrt{N}$) and the blue and green shading represents $\Delta$SFRs consistent with star-forming and green valley galaxies, respectively.}
        \label{props_vs_dsfr}
    \end{figure*}

    In Figure \ref{props_vs_dsfr}, we present the sSFR (left panel), $f_\text{gas}$ (centre panel), and SFE (right panel) as a function of $\Delta$SFR. In each panel, the median inner values are represented as red circles and the median outer values as black squares. For this test, we include upper/lower limits on the semi-resolved properties at face value when computing the medians. The shaded area around the lines represents the standard error in the median. Shaded blue and green regions reflect classifications of star-forming and green valley galaxies, respectively.

    Focusing first on the left panel of Figure \ref{props_vs_dsfr}, we find that both sSFR$_\text{inner}$ and sSFR$_\text{outer}$ are strongly correlated with global $\Delta$SFR. This is expected since both sSFR and $\Delta$SFR broadly capture the position of a galaxy relative to the SFMS. However, it is the difference (and lack thereof) between the inner and outer sSFR that is more interesting. Inspecting the continuous inner and outer sSFRs as they intersect with the SFMS region (blue shading) reveals that star-forming galaxies tend to have higher sSFR in their central regions, compared to their outer regions. Furthermore, the disparity between the sSFR$_\text{inner}$ and sSFR$_\text{outer}$ increases with increasing $\Delta$SFR. In other words, the scatter within the SFMS is largely dictated by the sSFR of the central region. Moreover, this widening disparity continues into the starburst region above the SFMS, indicating that global star formation enhancements are driven largely by the central region. In contrast, the median inner and outer sSFRs for galaxies below the SFMS are equal within errors, regardless of their $\Delta$SFR within the green valley. This sSFR parity indicates that global star-formation suppression is felt equally across inner and outer regions of suppressed galaxies.

    Shifting our attention to the central panel of Figure \ref{props_vs_dsfr}, we find that $f_\text{gas, inner}$ and $f_\text{gas, outer}$ are correlated with $\Delta$SFR, as expected given the KS relation. As with the sSFRs, the interest here lies in the difference (or lack thereof) between the inner and outer gas fractions as a function of $\Delta$SFR. For galaxies on the SFMS (blue shading), we find that the median inner and outer gas fractions are equal within error. Therefore, the molecular gas is evenly distributed within galaxies on the SFMS. Above the SFMS ($\Delta$SFR$>0.3$ dex), we find that the central gas fractions increase more than the gas fraction in the outer regions. Thus, \emph{global starbursts tend towards central gas enhancements} relative to their outer regions. Below the SFMS, we find that the $f_\text{gas, inner}$ decreases more than $f_\text{gas, outer}$. This suggests that \emph{global green valley galaxies tend towards central gas deficits} relative to their outer regions. We thus conclude that a galaxy's position above and below the SFMS is more sensitive to the central gas fractions than to the gas fraction in the outer regions of galaxies. 

    Finally, we present SFE$_\text{inner}$ and SFE$_\text{outer}$ as a function of $\Delta$SFR in the right panel of Figure \ref{props_vs_dsfr}. We find that in most $\Delta$SFR bins, the median SFE$_\text{inner}$ is higher than in the outer regions. However, below the SFMS (green shading) galaxies have approximately equally low SFE in the inner and outer regions. Above the SFMS, galaxies have $\thicksim$0.8 dex higher inner SFE relative to their outer SFE, regardless of $\Delta$SFR. The widening disparity between the inner and outer SFE happens entirely on the SFMS (blue shading). This demonstrates that the global star formation of galaxies on the SFMS is driven by the efficiency with which the central regions are forming stars.

    In summary, the semi-resolved properties of galaxies (Figure \ref{props_vs_dsfr}) have provided insight into the reasons for the scatter of the SFMS and the physics driving galaxies to depart from it. Galaxies on the main sequence have equal $f_\text{gas}$ in the inner and outer regions, but a variation in global SFR is driven by the degree of central sSFR enhancement brought about by higher central efficiency. Above the SFMS, the increase of $\Delta$SFR is driven both by an enhancement in inner $f_\text{gas}$ and higher central efficiency. However, below the SFMS, galaxies are equally low in inner and outer sSFR and with equally low SFEs paired with a systematic depletion in inner and outer molecular gas reservoirs, but more depleted in the central region by a factor of $\thicksim2$ at the bottom of the green valley.  

    \subsection{What dictates global star-formation?}
    \label{results_corr}

    We have found that the molecular gas reservoirs in the central regions are more strongly enhanced above the SFMS and more deficient below the SFMS, relative to the outer regions. The gas fraction of the outer regions scales with $\Delta$SFR, but less so than the inner regions. Thus, we now ask the question \emph{does central gas dictate global star formation?}

\begin{figure*}
        \centering
        \includegraphics[width=1\linewidth]{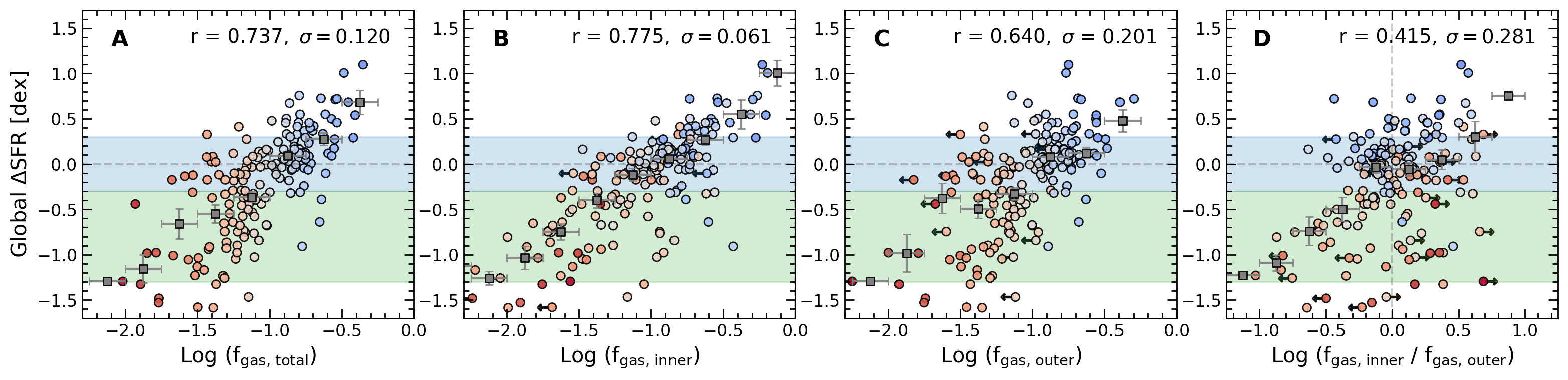}
        \caption{Global $\Delta$SFR as a function of $f_\text{gas, total}$ (panel A), $f_\text{gas, inner}$ (panel B), $f_\text{gas, outer}$ (panel C), and the ratio between the inner and outer gas fraction (panel D). The points are the CO-detected SALVAGE galaxies coloured by their $f_\text{gas, total}$. Gray squares are the median $\Delta$SFR in bins of the different gas fraction variables. The blue and green shading represents $\Delta$SFRs consistent with star-forming and green valley galaxies, respectively. The Pearson correlation coefficient (r) and the intrinsic scatter ($\sigma$) is in the top right corner of each panel. While the intrinsic scatter is a measure of the scatter after accounting of observational uncertainty, it may not be able to account for systematic biases, if present.} Global $\Delta$SFR is most strongly correlated with $f_\text{gas, inner}$ (and not $f_\text{gas, total}$) and the relationship between $f_\text{gas, inner}$ and $\Delta$SFR has the lowest intrinsic scatter.
        \label{dsfr_vs_fgas}
    \end{figure*}

    To answer this question, we consider $\Delta$SFR as our target variable and test how well it correlates with the total, inner, and outer gas fractions, as well as the ratio between the inner and outer gas fractions. In Figure \ref{dsfr_vs_fgas}, we present $\Delta$SFR as a function of $f_\text{gas, total}$ (panel A), $f_\text{gas, inner}$ (panel B), and $f_\text{gas, outer}$ (panel C), as well as the ratio between $f_\text{gas, inner}$ and $f_\text{gas, outer}$ (panel D). In each panel, the points are coloured by $f_\text{gas, total}$ (as quantified in panel A) and non-detections in the inner or outer regions are indicated by points with arrows. Gray squares show the median $\Delta$SFR in a given gas fraction bin; the x-errors represent the bin width and the y-errors represent the standard error in the median. 

    We measure the causal relationship between global $\Delta$SFR and the various gas fractions using the Pearson correlation coefficient (r), which quantifies the linear correlation between two variables, and the intrinsic scatter, accounting for errors on the SFR, M$_\star$ and M$_\text{mol}$ measurements with the \texttt{ltsfit} Python package \citep{Cappellari2013a}. Our measurements exclude any upper limit estimates (though our qualitative results do not change if we include upper limits at face value) and we disable outlier clipping in the \texttt{ltsfit} intrinsic scatter measurement. Panel B of Figure \ref{dsfr_vs_fgas} shows that $\Delta$SFR is more strongly correlated with $f_\text{gas, inner}$ than any of the other gas fraction variables we tested, both in terms of the highest correlation coefficient (r = 0.775) and the lowest intrinsic scatter of the relation ($\sigma = (0.061\pm0.035$) dex). 
    
    Panel A of Figure \ref{dsfr_vs_fgas} demonstrates that while the correlation between $f_\text{gas, total}$ and $\Delta$SFR is still strong (r = 0.737), it is not as strong as the correlation with $f_\text{gas, inner}$. The relationship between the global gas fraction and global $\Delta$SFR also has a larger intrinsic scatter of $\sigma = (0.120 \pm 0.039$) dex. The weaker correlation and larger scatter between the global gas fraction and global $\Delta$SFR arises from the even weaker correlation and large scatter in the relationship between the outer gas fraction and global $\Delta$SFR (r = 0.640, $\sigma = (0.201 \pm 0.037$) dex); panel C of Figure \ref{dsfr_vs_fgas} demonstrates that this is largely due to the lack of dynamic range in the outer gas fraction. For example, a majority of green valley galaxies span the same range of outer gas fraction as galaxies on the SFMS. Likewise, the strongest starburst galaxies have high $f_\text{gas, total}$ (dark blue) but relatively normal outer gas fractions, relative to those on the SFMS. The outer gas fraction is simply less responsive to $\Delta$SFR and vice versa. Though the outer region may be more likely to be affected by lower sensitivity and resolving out large scale flux, these qualitative results hold regardless of maximum resolvable scale and signal-to-noise cuts on the sample.

    Of the four gas fraction variables tested here, $\Delta$SFR exhibits the weakest correlation (r = 0.415) and the largest intrinsic scatter ($\sigma = (0.281 \pm 0.043$) dex) with the inner to outer gas fraction ratio. However, the fact that there is any correlation at all -- albeit a weak one -- is perhaps surprising. After all, the ratio between $f_\text{gas, inner}$ and$f_\text{gas, outer}$ tells you nothing about the total amount of molecular gas available to form stars, only about the relative distribution of the available gas. The correlation between $f_\text{gas, inner}/f_\text{gas, outer}$ and global $\Delta$SFR demonstrates that the movement of molecular gas throughout the galaxy towards or away from the centre has ramifications on the global star formation status of the galaxy.
\section{Discussion}
\label{discussion}

\subsection{The importance of the central region to quenching in the context of previous works}

SALVAGE combines the resolved CO maps and spectro-photometric catalogues of SDSS to independently resolve stellar mass, SFR, and molecular gas mass (and combinations of the three) in the inner and outer regions for 277 galaxies at $z\thicksim0.05$. This method allows us to bridge the gap between the detail (but small sample size) provided by fully resolved IFU surveys (e.g. ALMaQUEST, EDGE-CALIFA) and the statistically large (but unresolved) surveys (e.g. xCOLD GASS). \citet{Colombo25-iEDGE} recently released iEDGE, a database of optical-IFU spectroscopy combined with what we call in this work ``semi-resolved'' molecular gas observations. However, the semi-resolved perspective of iEDGE is subtly different from SALVAGE in that it relies on aperture corrections to infer the outer molecular gas mass, based on the central CO luminosity. The semi-resolved components are therefore not completely independent. Due to the resolution of the CO observations, iEDGE also has a larger central aperture for the inner region. Regardless, iEDGE is an excellent complementary approach to understanding the semi-resolved properties of quenching galaxies and we compare to both \citet{Colombo25-iEDGE} and \citet{Colombo25-quench} often in this section. 

Despite its relatively small area (1.3 kpc median inner radius set by the SDSS fibre), the gas fraction in the central region of massive ($M_{\star} \gtrsim 10^{10} M_\odot$) and CO-detected galaxies has the strongest correlation with the \emph{global} $\Delta$SFR, even more so than its global molecular gas mass. Hence, the central regions of galaxies have distinguished themselves as an especially crucial region in a galaxy, either driving the global change in SFR or highly sensitive to the same mechanisms that affect global SFR. Moreover, the claim that the central region plays an impactful role, fits into a framework of previous results from the radial SFR profiles in fully resolved IFU studies \citep[e.g.][]{Ellison18, Belfiore18}, wherein global SFR is most sensitive to the SFR in the central few kpc. In addition, studies such as \citet{Bluck14, Bluck20} have found central stellar density and central velocity dispersion to be the best predictor of galaxy quenching. 

On the SFMS, we find that the median inner and outer gas fractions are equal within error, regardless of the galaxy's position within the scatter of the main sequence (see Fig. \ref{props_vs_dsfr}). This is in agreement with \citet{Colombo25-iEDGE} and \citet{Colombo25-quench} who find equal gas fractions in the inner and outer regions of star-forming galaxies and studies of resolved gas fraction profiles \citep[e.g.][]{Pan24, Villanueva24}. Moreover, we find that the central SFE changes dramatically within the scatter of the main sequence; galaxies that scatter above the SFMS have a higher central SFE than in the outer regions. Generally speaking, we find that central SFE drives a galaxy's position on the SFMS. This is in agreement with the findings from \citet{Ellison20} showing that for galaxies with kpc-scale resolution in ALMaQUEST, SFE is a stronger predictor of distance from the SFMS than $f_\text{gas}$.

Above the main sequence, we find that galaxies have high sSFR$_\text{inner}$, driven by both a high $f_\text{gas, inner}$ and SFE$_\text{inner}$ (see Fig. \ref{props_vs_dsfr}). Therefore, starbursts are driven by both fuel and efficiency. Resolved studies of starbursting galaxies echo similar findings \citep{Ellison20-SBs}. In particular, mergers often have enhanced central gas fractions and lie above the SFMS  \citep{Garay-Solis23}. However, whether a merger-induced starburst is driven by fuel availability or efficiency changes from case-to-case \citep{Thorp22}. Indeed, 64 per cent (7/11) of the CO-detected galaxies in SALVAGE 0.5 dex or more above the SFMS appear to be post-mergers or interacting pairs, by visual inspection of their DECaLS images. Though our results agree that both enhanced central gas and SFE play a role in driving global $\Delta$SFR enhancements above the SFMS, the continued increase of $f_\text{gas, inner}$ and the plateau of SFE$_\text{inner}$ with $\Delta$SFR above the SFMS indicates the strength of the global starburst is controlled by enhanced central gas (see Fig. \ref{props_vs_dsfr}).

Below the main sequence, we find global suppression in SFE and common central gas depletion (Fig. \ref{props_vs_dsfr}). Several resolved works have focused on inferring whether quenching is driven by a lack of efficiency or a lack of available fuel \citep[e.g.][]{Lin22, Pan24, Villanueva24}. With resolved data, \citet{Lin22} computes a bulge/disk decomposition to probe whether a lack of fuel or efficiency in the central bulge of green valley galaxies is driving the suppressed SFR, finding that the answer changes from case-to-case. With a semi-resolved perspective, \citet{Colombo25-quench} find that quenching is driven by a lack of efficiency in the central regions, not a lack of available fuel. Our results point towards the exact opposite conclusion; we find quenching galaxies have lower central $f_\text{gas}$ than in the outer regions, and roughly constant SFE throughout.

\begin{figure}
        \centering
        \includegraphics[width=1\linewidth]{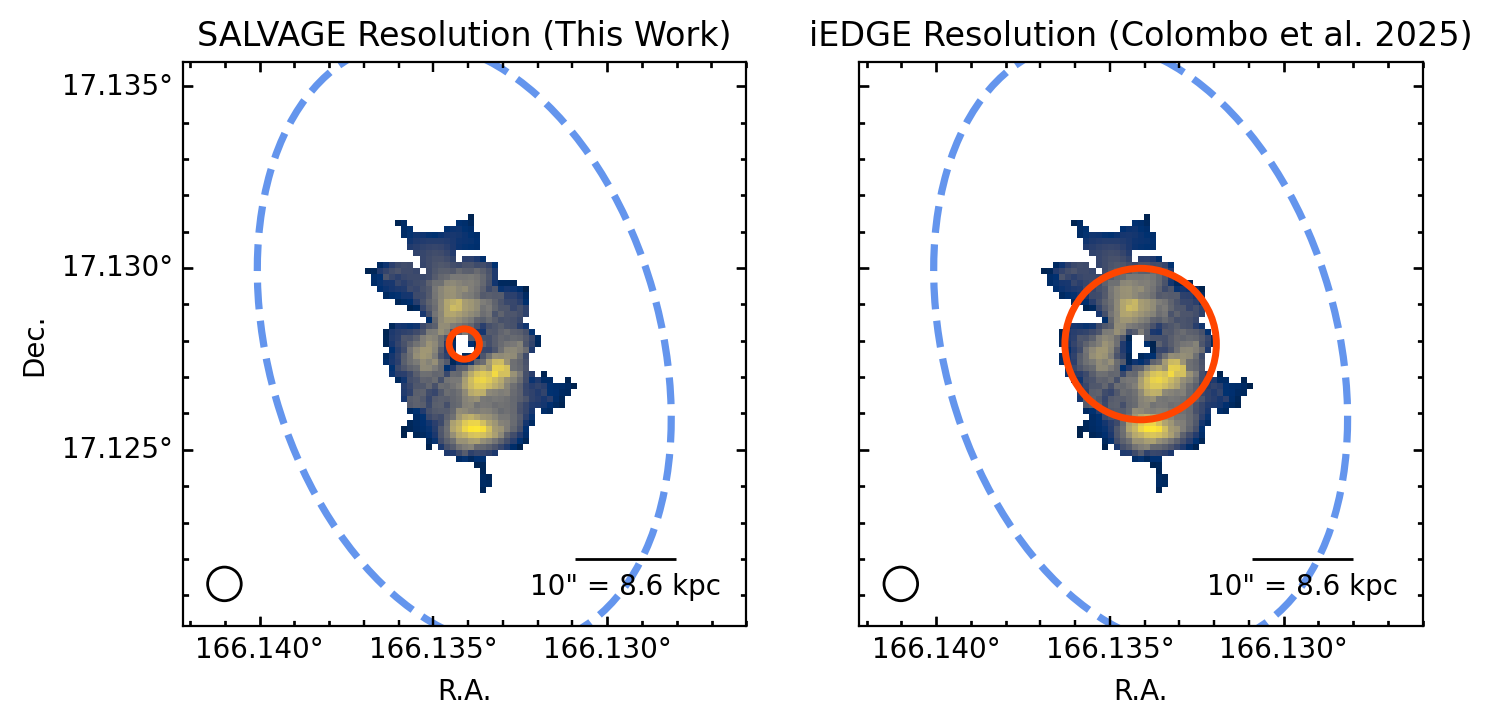}
        \caption{The ALMA CO line intensity map of SDSS Object 587742775089758339, a green valley galaxy with a $\Delta$SFR$= -0.73$ dex. In the left panel, the 3" SDSS fibre aperture is overlaid in red, representing the inner region of this SALVAGE target. The approximate extent of the SDSS photometric model in blue, representing the outer region. In the right panel, the Petrosian half light radius is overlaid in red, representing the approximate beam resolution of the iEDGE CO data \citep{Colombo25-iEDGE}. Along the bottom are the circularized beam of the ALMA data and a scale bar representing 10". \emph{If this galaxy were observed at the resolution used in iEDGE, the ``inner'' region would have a considerably higher molecular gas fraction and our conclusions would change as a result.}}
        \label{SALVAGEvsEDGE}
    \end{figure}

To unpack the difference between our conclusion (quenching is caused by a lack of central gas) and that of \citet{Colombo25-quench} (quenching is not caused by a lack of central gas, but rather a lack of efficiency), we present the CO intensity map of a SALVAGE galaxy found below the main sequence ($\Delta $SFR $= -0.73$ dex). This galaxy has a low inner gas fraction (log$(f_\text{gas, inner}) = -1.7$) compared to its outer gas fraction (log$(f_\text{gas, outer}) = -1.1$), both of which are representative of the median values for galaxies in the $\Delta $SFR range of $-1 < \Delta $SFR $<-0.5$ (see Fig. \ref{props_vs_dsfr}). In the left panel of Figure \ref{SALVAGEvsEDGE}, we plot our aperture definitions of inner and outer, according to the SDSS fibre aperture and radius of the SDSS modelMag photometry, respectively. In the right panel, we plot the iEDGE inner resolution element as the Petrosian half light radius as an approximate representation of the 1 $R_\text{e}$ quoted in \citet{Colombo25-iEDGE}. Ultimately, the region we refer to as the central regions differs from \citet{Colombo25-quench}. Our central regions are independently resolved and smaller ($\thicksim$1.3 kpc radius in this case) than those probed by \citet{Colombo25-quench} ($\thicksim 6.4$ kpc in this case). The central gas depletion in this galaxy (and many similar galaxies with suppressed SFR) occurs only within our smaller radius. Within the aperture to which iEDGE is sensitive, $f_\text{gas}$ would be normal or only marginally lower, which is common in quenching galaxies in iEDGE \citep{Colombo25-quench}. Furthermore, assuming the sSFR is constant across this galaxy (which true for the median sSFRs at this $\Delta$SFR, see Fig. \ref{props_vs_dsfr}), then an increase in molecular gas necessarily increases $f_\text{gas}$ and decreases SFE, leading to the opposite conclusions between this work and \citet{Colombo25-quench}. In summary, \textbf{whether quenching is caused by low efficiency or a lack of fuel in the inner/outer regions of galaxies depends on the resolution of the mm data}. The next paper in the series will demonstrate that molecular gas depletion within the central few kiloparsecs is a rather common occurrence in green valley galaxies (Wilkinson et al. in prep.).

\subsection{On the utility of the gas fraction ratio}

    The median values of $f_\text{gas, inner}$ and $f_\text{gas, outer}$ in Figure \ref{props_vs_dsfr} paint a picture in which many galaxies in the green valley have suppressed global SFR due to a lack of molecular gas in the inner regions. However, in panel D of Figure \ref{dsfr_vs_fgas}, we found that $\Delta$SFR was correlated with the ratio between the inner and outer gas fractions, but with significant scatter in the green valley. We revisit this result now to discuss the possibility of using the distribution of molecular gas to distinguish between different quenching mechanisms. In particular, we explore mergers and strong bars identified by GalaxyZoo morphological classification of DECaLS imaging \citep{Walmsley23}, as well as AGN identified as being above the BPT diagram with a S/N$>3$ on all four emission lines and EW(H$\alpha) < -3$ Å. 

    \begin{figure*}
        \centering
        \includegraphics[width=1\linewidth]{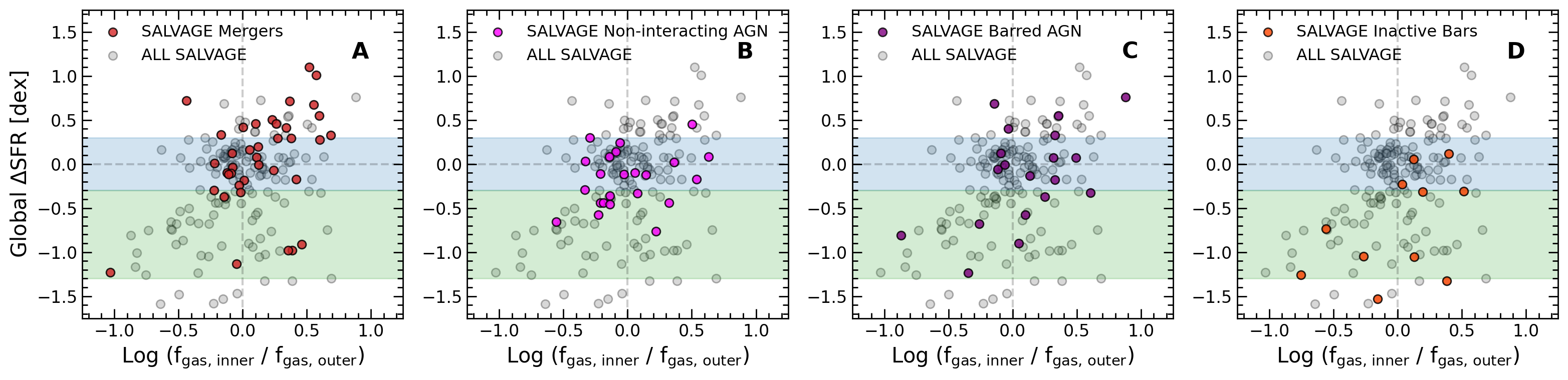}
        \caption{Proposed evolution of $\Delta$SFR and the inner/outer gas fraction ratio for quenching induced by mergers, AGN, barred AGN, and morphological quenching. Coloured points refer to SALVAGE targets with a given identifiable quenching mechanism present, as described in the text. Gray points refer to all other CO-detected targets in SALVAGE. The blue and green shading represents $\Delta$SFRs consistent with star-forming and green valley galaxies, respectively.}
        \label{mechanisms}
    \end{figure*}
    
    In Figure \ref{mechanisms}, we now explore the position of various specific populations or events on the distribution of molecular gas and its relationship to $\Delta$SFR. In panel A of Figure \ref{mechanisms}, we first consider mergers. Simulations and observations show evidence that galaxy mergers may induce gravitational torques that destabilize gas, causing inflows towards the centre of the galaxy \citep{Toomre72, Barnes91, Scudder12, Blumenthal18}. The inflowing gas often coincides with a central starburst \citep{Moreno15, Hani2020, Shah22}. Furthermore, observations of post-mergers have shown that they can rapidly quench global SFR within a few hundred Myr \citep{Ellison22, Ellison24-psbmergers}. Resolved observations of the molecular gas in post-starburst galaxies, often post-mergers themselves, tend to host highly compact molecular gas reservoirs \citep{Smercina22, Otter22, Wilkinson22}. Therefore, our expectation is that mergers would move up and to the right as merger-triggered inflow induces a starburst, and then down as the star formation ceases, leaving behind a compact molecular gas reservoir (if any molecular gas is detected at all). In panel A of Figure \ref{mechanisms}, we highlight the mergers (red points) using the DECaLS GalaxyZoo classifications (supplemented by visual classification by the authors where GalaxyZoo missed obvious merger features) and find that they tend to have normal or enhanced central molecular gas reservoirs. The post-mergers in the green valley with $\log{f_\text{gas, inner} / f_\text{gas, outer}} > 0$ are indeed post-starburst galaxies as classified by the methods in \citet{Wilkinson22}. Therefore, the post-merger/post-starburst rapid quenching route is contributing to a population of galaxies with centrally concentrated molecular gas reservoirs in the green valley.

    Next, we consider the effect of AGN, which theory predicts can produce radiative and mechanical feedback that affects the surrounding gas \citep{Hopkins08, Hopkins09-effectofgas}. Indeed, simulations have shown that when AGN feedback is turned on, molecular gas reservoirs are centrally depleted relative to their outer regions \citep{Terrazas20, Appleby20}. However, observations of low-redshift AGN often find normal or enhanced molecular gas reservoirs \citep[e.g.][]{Koss21, RA22, Salvestrini22, Molina23}. Recent studies using fully resolved observations have shown some signal of depletion in the central regions of AGN \citep{Ellison21, AA23, GB24}. If AGN deplete molecular gas starting from their central regions and lead to global galaxy quenching, we would expect galaxies to be towards the lower left of the panel. Highlighting non-merger, non-barred AGN in panel B of Figure \ref{mechanisms}, we find that some (but not all) AGN follow this prediction. We note that other galaxies in the lower left panel (i.e. quenching and centrally depleted in molecular gas) may be AGN that have recently shut off and therefore cannot be detected as such due to a lack of emission lines (or a lack of central gas to ionize). We explore the nature of these centrally gas deficient objects and if they are indeed caused by AGN in the next paper in the series (Wilkinson et al. in prep.).

    Bars can form secularly in rotating disks from differential density waves and have been shown to trigger inflow of gas from the outer regions of a galaxy towards the centre \citep{Athanassoula92}. The high central gas densities induced by bars have been shown to trigger AGN \citep{Alonso24, Garland24, Marels25}, which may, in turn, remove central gas via feedback. The competing effects of gas inflow triggered by the bar and potential feedback from the AGN agree with the location of Barred-AGN in panel C of Figure \ref{mechanisms}: barred-AGN show a large scatter including centrally enhanced, normal, and deficient molecular gas. 

    Lastly, the formation of bars and bulges has also been shown to induce morphological quenching by stabilizing the gas against collapse into stars \citep{Khoperskov18}. In this scenario, the inner and outer gas fractions remain relatively unchanged, but global SFR decreases. Therefore, we predict that galaxies experiencing morphological quenching would move straight down in Figure \ref{mechanisms}. Non-merger, non-AGN barred galaxies show a large scatter in panel D of Figure \ref{mechanisms}, but do exhibit broad agreement with this prediction. 

    Mapping experiments of molecular gas in early-type galaxies show that when molecular gas remains, it can take on several different morphologies including disks, rings, and unresolved/compact central gas \citep{Alatalo13}. The work presented here provides a glimpse into the prospect of diagnosing quenching mechanisms on the basis of the molecular gas distribution. The path towards reliably connecting CO distributions to quenching mechanisms includes mapping CO in a large and diverse sample of galaxies in combination with generating realistic gas maps from cosmological simulations \citep[e.g.][]{Lagos15, Perron-Cormier25} where quenching mechanisms can be traced backwards in time \citep[e.g.][]{Davis19, Pawlik19}.

\subsection{On the assumption of a constant $\alpha_{CO}$}
\label{discussion_alpha}

The conversion from $L_\text{CO}$ to $M_\text{mol}$ is sensitive to the emissivity of the CO, which has been shown to change in different locations within the galaxy, and vary from galaxy to galaxy \citep[e.g.][]{Narayanan11, Sandstrom13, He24, Chiang24}. For the CO~($1-0$) line in particular (as opposed to higher-J transition lines), $\alpha_\text{CO}$ is most affected by the local gas phase metallicity and CO line width \citep{Teng23, Teng24, Chiang24}. Furthermore, due to variations of these effects throughout a galaxy, $\alpha_\text{CO}$ has been shown to deviate from the Milky Way conversion radially; $\alpha_\text{CO}$ tends to lower values in the central regions of galaxies \citep[e.g.][]{Sandstrom13, Chiang24}. By assuming a constant Milky Way conversion throughout the galaxy, could we have overestimated the central molecular gas mass and therefore the relative contribution of gas availability versus star formation efficiency in the central regions to global SFR?

Following the work by \citet{Chiang24}, \citet{Schinnerer24} recommend an $\alpha_\text{CO}$ that varies from the constant Milky Way conversion factor of $\alpha_\text{CO} = 4.35$ M$_\odot$ (K km s$^{-1}$ pc$^{2}$)$^{-1}$ by a ``CO dark'' term that depends on metallicity and a ``starburst emissivity'' term that depends on the stellar mass surface density ($\Sigma_\star$) as a proxy for CO linewidth, recent starburst, or an otherwise turbulent ISM that would decrease the opacity of CO. The CO dark term, $f(Z)$, is defined as 

\begin{equation}
    f(Z) = \left(\frac{Z}{Z_\odot}\right)^{-1.5}, 
\end{equation}

\noindent where $Z$ is the gas-phase metallicity and $Z_\odot = 0.0134$ \citep{Asplund09, Chiang24}. The starburst emissivity term, $g(\Sigma_\star)$, is defined as 

\begin{equation}
    g(\Sigma_\star) = \left(\frac{\text{max}(\Sigma_\star \text{, 100 M}_\odot\text{ pc}^{-2})}{\text{100 M}_\odot\text{ pc}^{-2}}\right)^{-0.25}.
\end{equation}

Incorporating these terms together, the conversion factor recommended by \citet{Schinnerer24} becomes

\begin{equation}
    \alpha_\text{CO}(Z, \Sigma_\star) = \alpha_\text{CO,MW}f(Z)g(\Sigma_\star), 
\end{equation}

\noindent where $\alpha_\text{CO, MW}$ is 4.35 M$_\odot$ (K km s$^{-1}$ pc$^{2}$)$^{-1}$.

    \begin{figure}
        \centering
        \includegraphics[width=1\linewidth]{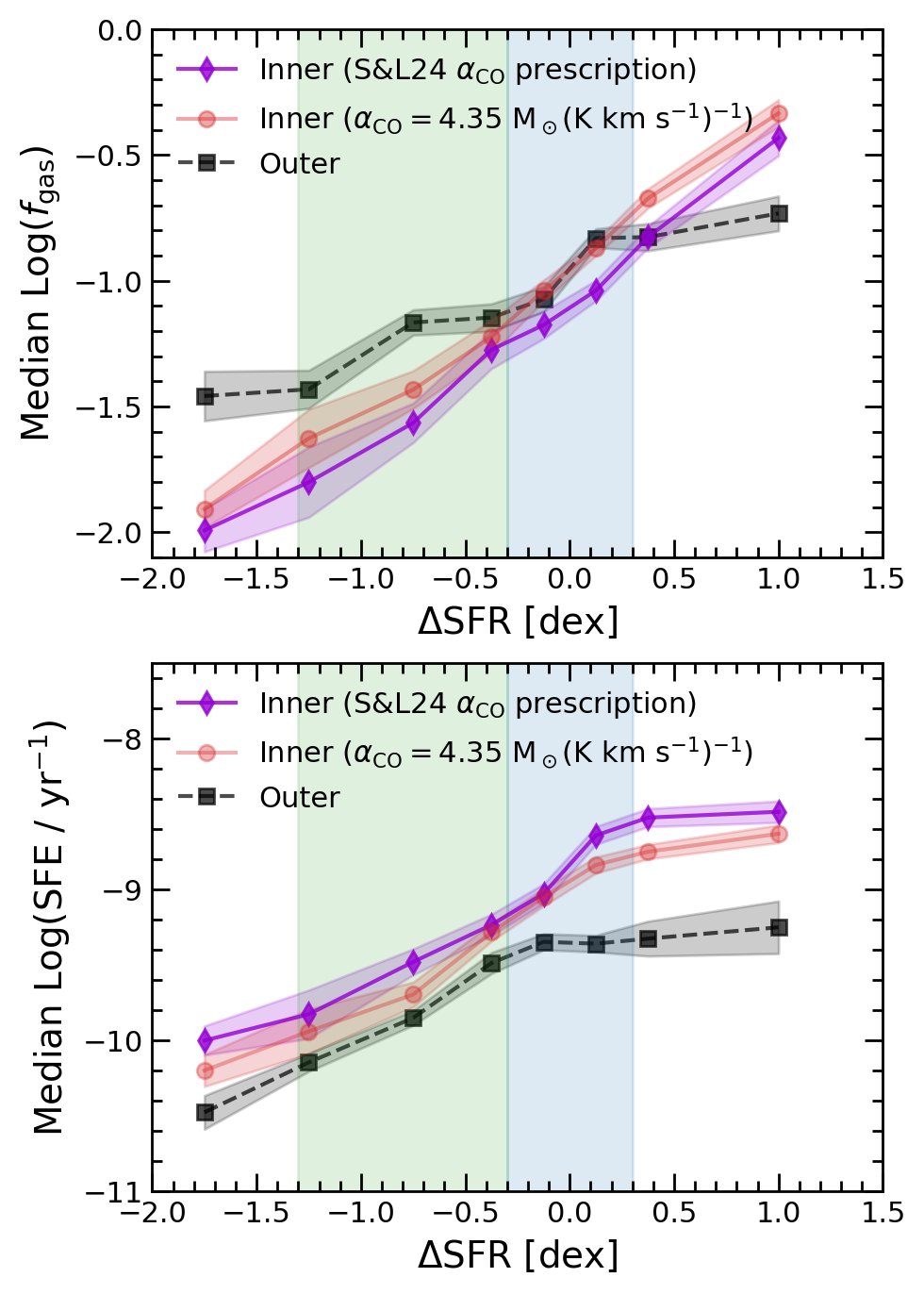}
        \caption{The median $f_\text{gas}$ (top panel), and SFE (bottom panel) as computed with an $\alpha_\text{CO} = 4.35$ (K km s$^{-1}$ pc$^{2}$)$^{-1}$ for the inner (red circles) and outer (black squares) regions of SALVAGE targets as a function of $\Delta$SFR (i.e. the same as in Figure \ref{props_vs_dsfr}). Purple diamonds represent the median $f_\text{gas, inner}$ and SFE$_\text{inner}$ as computed using an alternative $\alpha_\text{CO}$ conversion factor suggested by \citet{Schinnerer24}. Each $\Delta$SFR bin spans 0.5 dex except for the region $-0.5 < \Delta$SFR$< 0.5$ where there are enough targets to reduce bins to 0.25 dex. The shaded area around the lines represent the standard error in the median ($\sigma / \sqrt{N}$) and the blue and green shading represents $\Delta$SFRs consistent with star-forming and green valley galaxies, respectively. \emph{Our qualitative results hold regardless of changes to the assumed $\alpha_\text{CO}$ conversion factor.}}
        \label{alpha_co}
    \end{figure}

To test whether our qualitative results are significantly affected by our choice of $\alpha_\text{CO}$, we recompute $f_\text{gas, inner}$ and SFE$_\text{inner}$ for every galaxy in SALVAGE using the \citet{Schinnerer24} recommended prescription. We measure the gas-phase metallicity within the SDSS fibre from the [OIII] and [NII] emission lines following \citet{Curti17} and an assumed $12 + \log\left({\text{O}/\text{H}}\right)_\odot$ of 8.69 \citep{Asplund09, Chiang24}. When the [OIII] and [NII] emission lines are not present, we assume solar metallicity such that $f(Z)=1$. $\Sigma_{\star\text{,inner}}$ is computed as M$_\star,\text{inner}$ divided by the area covered by the SDSS fibre in pc, inclination-corrected by $\cos{i}$. We do not apply this prescription to the outer regions for several reasons. First, our results are most sensitive to the inner regions, thus it is most important that we rigorously check these results. Second, the ``edge'' of the outer region is not well-defined for each galaxy and can cause significant uncertainty in $\Sigma_{\star,\text{outer}}$. Lastly, the metallicity measurements from SDSS are based on emission lines that apply to the area within the fibre only. 

In Figure \ref{alpha_co}, we present the semi-resolved measurements of $f_\text{gas}$ and SFE as a function of $\Delta$SFR (the same as the centre and right panels of Figure \ref{props_vs_dsfr}), but with an alternative $\alpha_\text{CO}$ prescription from \citet{Schinnerer24} applied to the inner region. We find that our main qualitative results hold: SFE$_\text{inner}$ trends strongly with $\Delta$SFR for galaxies on the SFMS, but plateaus above the main sequence, $f_\text{gas, inner}$ is enhanced above the SFMS, and depleted below the SFMS, relative to the outer regions.
\section{Summary}
\label{Summary}

    Starting with the main galaxy sample from SDSS DR7, we conducted a comprehensive search for, and data reduction of, CO~(1-0) data from the ALMA Science Archive. The result is a sample of 277 galaxies at $z\thicksim0.05$ with ``semi-resolved'' inner and outer stellar mass, SFR, and molecular gas mass measurements. The combination of both optical data from SDSS to study the stellar populations and millimetre data to study the molecular gas allows us to understand the interplay between the star-forming material (or lack thereof) and the stars they form (or cease to form). We leverage this dataset to provide a semi-resolved perspective of established global scaling relations. Our main conclusions are as follows:

    \begin{itemize}

        \item \textbf{The global SFMS is sensitive to where the star-formation occurs within the galaxy, but the global MGMS is not sensitive to where the molecular gas is within the galaxy.} Galaxies above the SFMS tend to have high central sSFR (relative to their outer regions) and green valley galaxies below the SFMS tend to have low central sSFR. No such trend exists for the MGMS, indicating the distribution of molecular gas within a galaxy has no bearing on its position on the MGMS. The same is true for the KS relation, but there is a signal that some galaxies below are caused by inefficient central regions (see Figure \ref{scalinglaws}).
        \smallskip

        \item \textbf{A galaxy's position within the SFMS is largely dictated by the star-formation efficiency in the central few kpc.} Our interpretation is that this contributes to the scatter of the SFMS (see right panel of Figure \ref{props_vs_dsfr}).
        \smallskip

        \item \textbf{Central gas dictates the global star-forming status, more so than the total molecular gas reservoir.} The inner gas fraction is more sensitive to changes in global $\Delta$SFR than the outer gas fraction (see centre panel of Figure \ref{props_vs_dsfr}). Furthermore, global $\Delta$SFR shows a stronger correlation with lower instrinsic scatter with $f_\text{gas, inner}$  (r = 0.775, $\sigma = (0.061\pm0.035$) dex) than with $f_\text{gas, total}$ (r = 0.737, $\sigma = (0.120 \pm 0.039$) dex; see Figure \ref{dsfr_vs_fgas}).
        \smallskip
        
    \end{itemize}

    SALVAGE underscores the importance of leveraging archival data, the utility of semi-resolved data for understanding galaxy evolution, and the need for large samples of resolved CO observations to capture the diversity of the galaxy population. With this large sample of semi-resolved galaxies, we have found the central region of the galaxy to play a particularly significant role in galaxy evolution. However, when placing this work (and other works exploring the importance of the central region) in the context of the literature, we urge readers to consider the physical size of the ``central'' region, as we have shown conclusions may vary due to the resolution of the data and the definition of the central region.

\section*{Acknowledgements}
\label{acknowledgements}

We respectfully acknowledge the L\textschwa\textvbaraccent {k}$^{\rm w}$\textschwa\ng{}\textschwa n Peoples on whose traditional territory the University of Victoria stands and the Songhees, Esquimalt and $\underline{\text{W}}\acute{\text{S}}$ANE$\acute{\text{C}}$ peoples whose relationships with the land continue to this day. As we explore the shared sky, we acknowledge our responsibilities to honour those who were here before us, and their continuing relationships to these lands. We strive for respectful relationships and partnerships with all the peoples of these lands as we move forward together towards reconciliation and decolonization. 

SW would like to thank Simon Smith, Shoshannah Byrne-Mamahit, Ben Rasmussen, Dave Patton, and Dario Colombo for their engaging scientific discussions that improved the quality of this work. SW would like to acknowledge the support received from Helen Kirk who maintained the necessary software on the CANFAR Science Platform and went above and beyond to assist with the ALMA data reduction. We also gratefully acknowledge the CADC/CANFAR staff whose hard work to build and maintain the CANFAR Science Platform (and its affiliated resources) made this work possible. We also thank the ALMA Help Desk staff for their assistance in optimizing the archival query services, for providing calibrated measurement sets for observations for which we could not restore pipeline calibrations, and for their generous support.

SW gratefully acknowledges the support from the Natural Sciences and Engineering Council of Canada (NSERC) as part of their graduate fellowship program. SLE gratefully acknowledges the receipt of NSERC Discovery Grants. Cette recherche a été financée par le Conseil de recherches en sciences naturelles et en génie du Canada (CRSNG).

SW and CC acknowledge support from the ESA Science Faculty Visitor Funding, reference ESA-SCI-E-LE-094.

The authors acknowledge the use of the Canadian Advanced Network for Astronomy Research (CANFAR) Science Platform. Our work used the facilities of the Canadian Astronomy Data Center, operated by the National Research Council of Canada with the support of the Canadian Space Agency, and CANFAR, a consortium that serves the data-intensive storage, access, and processing needs of university groups and centers engaged in astronomy research.

This research used the Canadian Advanced Network For Astronomy Research (CANFAR) operated in partnership by the Canadian Astronomy Data Centre and The Digital Research Alliance of Canada with support from the National Research Council of Canada the Canadian Space Agency, CANARIE and the Canadian Foundation for Innovation.

Funding for the SDSS and SDSS-II has been provided by the Alfred P. Sloan Foundation, the Participating Institutions, the National Science Foundation, the U.S. Department of Energy, the National Aeronautics and Space Administration, the Japanese Monbukagakusho, the Max Planck Society, and the Higher Education Funding Council for England. The SDSS Web Site is http://www.sdss.org/.

The SDSS is managed by the Astrophysical Research Consortium for the Participating Institutions. The Participating Institutions are the American Museum of Natural History, Astrophysical Institute Potsdam, University of Basel, University of Cambridge, Case Western Reserve University, University of Chicago, Drexel University, Fermilab, the Institute for Advanced Study, the Japan Participation Group, Johns Hopkins University, the Joint Institute for Nuclear Astrophysics, the Kavli Institute for Particle Astrophysics and Cosmology, the Korean Scientist Group, the Chinese Academy of Sciences (LAMOST), Los Alamos National Laboratory, the Max-Planck-Institute for Astronomy (MPIA), the Max-Planck-Institute for Astrophysics (MPA), New Mexico State University, Ohio State University, University of Pittsburgh, University of Portsmouth, Princeton University, the United States Naval Observatory, and the University of Washington.

This paper makes use of the following ALMA data: \\
ADS/JAO.ALMA\#2011.0.00374.S, ADS/JAO.ALMA\#2012.1.00539.S, ADS/JAO.ALMA\#2013.1.00058.S, ADS/JAO.ALMA\#2013.1.00096.S, ADS/JAO.ALMA\#2013.1.00115.S, ADS/JAO.ALMA\#2013.1.00530.S, ADS/JAO.ALMA\#2013.1.01383.S, ADS/JAO.ALMA\#2015.1.00320.S, ADS/JAO.ALMA\#2015.1.00389.S, ADS/JAO.ALMA\#2015.1.00405.S, ADS/JAO.ALMA\#2015.1.00587.S, ADS/JAO.ALMA\#2015.1.00820.S, ADS/JAO.ALMA\#2015.1.01012.S, ADS/JAO.ALMA\#2015.1.01120.S, ADS/JAO.ALMA\#2015.1.01225.S, ADS/JAO.ALMA\#2016.1.00177.S, ADS/JAO.ALMA\#2016.1.00329.S, ADS/JAO.ALMA\#2016.1.00852.S, ADS/JAO.ALMA\#2016.1.00948.S, ADS/JAO.ALMA\#2016.1.01172.S, ADS/JAO.ALMA\#2016.1.01265.S, ADS/JAO.ALMA\#2016.1.01269.S, ADS/JAO.ALMA\#2017.1.00025.S, ADS/JAO.ALMA\#2017.1.00496.S, ADS/JAO.ALMA\#2017.1.00601.S, ADS/JAO.ALMA\#2017.1.00629.S, ADS/JAO.ALMA\#2017.1.01093.S, ADS/JAO.ALMA\#2017.1.01727.S, ADS/JAO.ALMA\#2018.1.00541.S, ADS/JAO.ALMA\#2018.1.00558.S, ADS/JAO.ALMA\#2018.1.00940.S, ADS/JAO.ALMA\#2018.1.01852.S, ADS/JAO.ALMA\#2019.1.00260.S, ADS/JAO.ALMA\#2019.1.00597.S, ADS/JAO.ALMA\#2019.1.01136.S, ADS/JAO.ALMA\#2019.1.01757.S, ADS/JAO.ALMA\#2021.1.00094.S, ADS/JAO.ALMA\#2021.1.00602.S, ADS/JAO.ALMA\#2021.1.01089.S, ADS/JAO.ALMA\#2022.1.00482.S. ALMA is a partnership of ESO (representing its member states), NSF (USA) and NINS (Japan), together with NRC (Canada), NSTC and ASIAA (Taiwan), and KASI (Republic of Korea), in cooperation with the Republic of Chile. The Joint ALMA Observatory is operated by ESO, AUI/NRAO and NAOJ. The National Radio Astronomy Observatory is a facility of the National Science Foundation operated under cooperative agreement by Associated Universities, Inc.

The Legacy Surveys consist of three individual and complementary projects: the Dark Energy Camera Legacy Survey (DECaLS; Proposal ID \#2014B-0404; PIs: David Schlegel and Arjun Dey), the Beijing-Arizona Sky Survey (BASS; NOAO Prop. ID \#2015A-0801; PIs: Zhou Xu and Xiaohui Fan), and the Mayall z-band Legacy Survey (MzLS; Prop. ID \#2016A-0453; PI: Arjun Dey). DECaLS, BASS and MzLS together include data obtained, respectively, at the Blanco telescope, Cerro Tololo Inter-American Observatory, NSF’s NOIRLab; the Bok telescope, Steward Observatory, University of Arizona; and the Mayall telescope, Kitt Peak National Observatory, NOIRLab. Pipeline processing and analyses of the data were supported by NOIRLab and the Lawrence Berkeley National Laboratory (LBNL). The Legacy Surveys project is honored to be permitted to conduct astronomical research on Iolkam Du’ag (Kitt Peak), a mountain with particular significance to the Tohono O’odham Nation.

NOIRLab is operated by the Association of Universities for Research in Astronomy (AURA) under a cooperative agreement with the National Science Foundation. LBNL is managed by the Regents of the University of California under contract to the U.S. Department of Energy.

This project used data obtained with the Dark Energy Camera (DECam), which was constructed by the Dark Energy Survey (DES) collaboration. Funding for the DES Projects has been provided by the U.S. Department of Energy, the U.S. National Science Foundation, the Ministry of Science and Education of Spain, the Science and Technology Facilities Council of the United Kingdom, the Higher Education Funding Council for England, the National Center for Supercomputing Applications at the University of Illinois at Urbana-Champaign, the Kavli Institute of Cosmological Physics at the University of Chicago, Center for Cosmology and Astro-Particle Physics at the Ohio State University, the Mitchell Institute for Fundamental Physics and Astronomy at Texas A\&M University, Financiadora de Estudos e Projetos, Fundacao Carlos Chagas Filho de Amparo, Financiadora de Estudos e Projetos, Fundacao Carlos Chagas Filho de Amparo a Pesquisa do Estado do Rio de Janeiro, Conselho Nacional de Desenvolvimento Cientifico e Tecnologico and the Ministerio da Ciencia, Tecnologia e Inovacao, the Deutsche Forschungsgemeinschaft and the Collaborating Institutions in the Dark Energy Survey. The Collaborating Institutions are Argonne National Laboratory, the University of California at Santa Cruz, the University of Cambridge, Centro de Investigaciones Energeticas, Medioambientales y Tecnologicas-Madrid, the University of Chicago, University College London, the DES-Brazil Consortium, the University of Edinburgh, the Eidgenossische Technische Hochschule (ETH) Zurich, Fermi National Accelerator Laboratory, the University of Illinois at Urbana-Champaign, the Institut de Ciencies de l’Espai (IEEC/CSIC), the Institut de Fisica d’Altes Energies, Lawrence Berkeley National Laboratory, the Ludwig Maximilians Universitat Munchen and the associated Excellence Cluster Universe, the University of Michigan, NSF’s NOIRLab, the University of Nottingham, the Ohio State University, the University of Pennsylvania, the University of Portsmouth, SLAC National Accelerator Laboratory, Stanford University, the University of Sussex, and Texas A\&M University.

BASS is a key project of the Telescope Access Program (TAP), which has been funded by the National Astronomical Observatories of China, the Chinese Academy of Sciences (the Strategic Priority Research Program “The Emergence of Cosmological Structures” Grant \# XDB09000000), and the Special Fund for Astronomy from the Ministry of Finance. The BASS is also supported by the External Cooperation Program of Chinese Academy of Sciences (Grant \# 114A11KYSB20160057), and Chinese National Natural Science Foundation (Grant \# 12120101003, \# 11433005).

The Legacy Survey team makes use of data products from the Near-Earth Object Wide-field Infrared Survey Explorer (NEOWISE), which is a project of the Jet Propulsion Laboratory/California Institute of Technology. NEOWISE is funded by the National Aeronautics and Space Administration.

The Legacy Surveys imaging of the DESI footprint is supported by the Director, Office of Science, Office of High Energy Physics of the U.S. Department of Energy under Contract No. DE-AC02-05CH1123, by the National Energy Research Scientific Computing Center, a DOE Office of Science User Facility under the same contract; and by the U.S. National Science Foundation, Division of Astronomical Sciences under Contract No. AST-0950945 to NOAO.

\section*{Data Availability}

The reduced ALMA cubes, moment maps, and higher-order products that were produced by the SALVAGE project are available at \url{https://www.canfar.net/storage/vault/list/AstroDataCitationDOI/CISTI.CANFAR/25.0077/data}. The PHANGS-ALMA imaging pipeline is available at \url{https://github.com/akleroy/phangs\_imaging\_scripts}, and the SALVAGE implementation of this pipeline, along with tutorials for interacting with the data, is available at \url{https://github.com/sj-wilkinson/SALVAGE}.

\bibliographystyle{mnras}
\bibliography{Bibliography}

\appendix
\newpage
\section{A comparison between different Star formation rate measurements}
\label{GSWLC}

The global SFRs used in SALVAGE come from the MPA/JHU catalogue. While this catalogue has been used widely, its mixed approach -- using spectroscopic methods for the inner region and photometric methods for the outer region -- could affect our results. In this Appendix, we compare total SFRs from MPA/JHU, which mixes spectroscopy and photometry data, with total SFRs from the GALEX-Sloan-WISE Legacy Catalogue (GSWLC), which uses only photometry data, to demonstrate that mixing SFR estimation methods is not driving our central conclusions. 

\subsection{Comparing MPA-JHU SFRs to GSWLC}

To address the potential for errors introduced by mixing the inner (spectroscopic) and outer (photometric) components to obtain the total SFR, we compare the total SFRs used in this work with photometric-only total SFRs from the GSWLC \citep{Salim16}. GSWLC uses methods similar to the photometric fits in the outer region, but for the total photometry. GSWLC uses SDSS \emph{ugriz} photometry, supplemented with mid-IR data from WISE and UV data from GALEX. GSWLC has several sub-catalogues depending on the depth of the UV data: shallow “all-sky” UV data (GSWLC-A), medium depth and coverage (GSWLC-M) and deep UV data (GSWLC-D). There are also two versions for each sub-catalogue, with the difference between versions 1 and 2 being the addition of free parameters used to describe the dust attenuation curve. For comparing with MPA/JHU, we will use GSWLC-A1, which is closest to MPA/JHU due to its lack of deep UV data. By comparing to this catalogue, we can see if the inclusion of a second (spectroscopic) SFR method introduces a systematic bias in the total SFR. We tested all variations of the GSWLC and our conclusions do not change.

For the purposes of testing the impact of SFR estimation methods, we select all galaxies in common between the main galaxy sample of SDSS DR7 with MPA/JHU SFR and M$_{\star\text{, total}}$ measurements and galaxies in the GSWLC-A1 catalogue with reliable SFR and M$_{\star\text{, total}}$ measurements (i.e. not equal to $-99$). To assess a parameter space comparable to galaxies in SALVAGE, we further restrict this overlap to only galaxies with $\log(\text{M}_\text{MPA/JHU}/\text{M}_\odot)>9.5$ and $0.01<z<0.015$. There are 318,620 galaxies in common that meet these criteria. Since GSWLC assumes a \citet{Chabrier03} initial mass function (IMF) and MPA/JHU assumes a \citet{Kroupa01} IMF, we add 0.025 dex to the SFR and stellar mass measurements in GSWLC to be directly comparable to MPA/JHU \citep{Salim07}.

\begin{figure}
        \centering
        \includegraphics[width=1\linewidth]{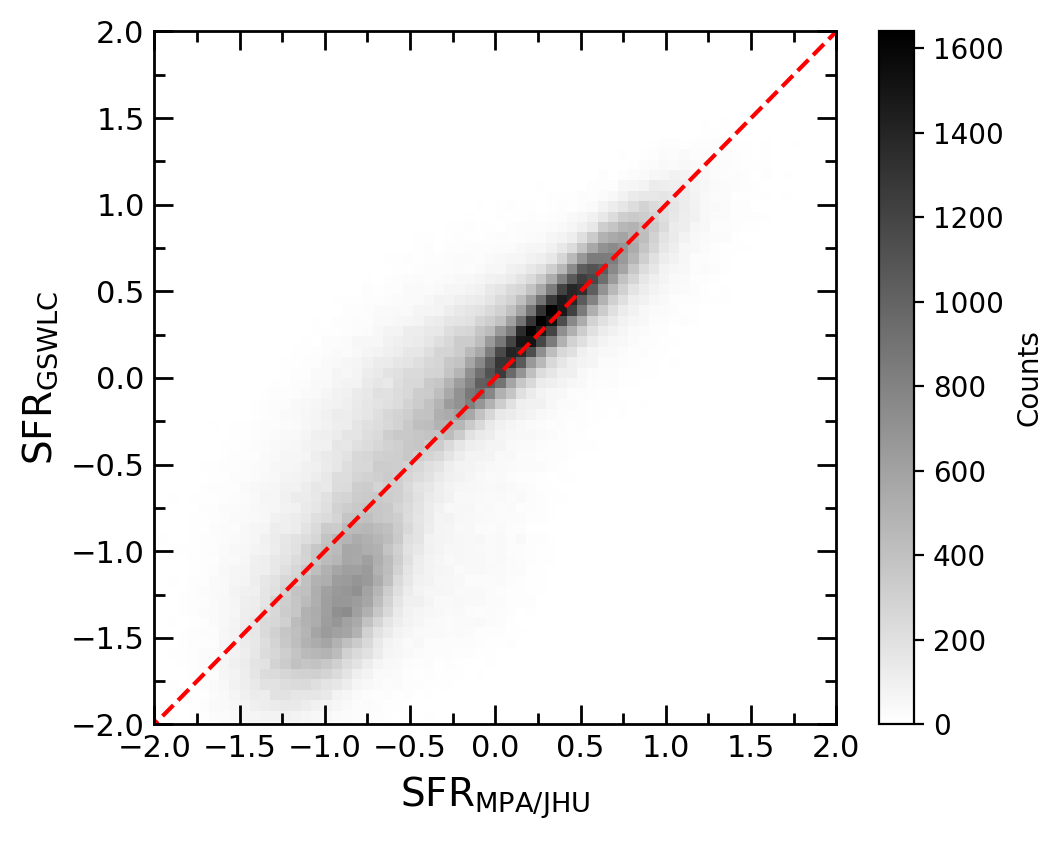}
        \caption{The total SFR measured with photometry only in GSWLC compared to the SFR measured with spectroscopy (inner) and photometry (outer). The black 2d histogram represents the number density of the comparison. The red dashed line represents the 1-to-1 line. Generally speaking, the SFRs for star forming galaxies agree, intermediate SFRs tend to be higher in GSWLC, and low SFRs are lower in GSWLC.}
        \label{mpa_vs_gsw}
    \end{figure}

In Figure \ref{mpa_vs_gsw}, we show a direct comparison between the SFR measurements from GSWLC and MPA/JHU; \citet{Salim16} conducted a detailed comparison of these two catalogues, so here we reproduce this comparison for internal consistency. The median absolute difference between the total SFRs in the two surveys is 0.21 dex, mostly driven by the offset between the red sequence galaxies, which are known to be unconstrained upper limits in MPA/JHU. Looking specifically at the star-forming galaxies (sSFR $>10^{-10.5} \text{ yr}^{-1}$), the median absolute difference is 0.11 dex and 83\% are equal within 0.25 dex, the typical $1\sigma$ uncertainty of the MPA/JHU SFR measurements. At intermediate SFRs, the GSWLC SFRs tend to be higher than those in the MPA/JHU catalogue. The galaxies with higher SFRs in GSWLC tend to have strong colour gradients (red centers with blue outskirts) and emit brightly in the mid-IR (WISE band 3); thus, we suspect MPA/JHU may be missing star-formation due to dust-obscuration, but do not rule out overestimation of the SFR in GSWLC due to IR contributions of the central aging stellar population or AGN. A complete assessment of the underlying causes of differences between the two catalogues is beyond the scope of this work. \emph{Since the SFRs of well-constrained star-forming galaxies are largely consistent, we conclude that mixing spectroscopic and photometric estimates of SFR does not lead to systematic differences.}

\subsection{Changing global SFR methods does not impact our results}

In the previous subsection, we have shown that while the global SFR of star-forming galaxies are typically equal within error, galaxies with intermediate SFRs tend to be higher in GSWLC by up to $\thicksim0.5$ dex. GSWLC uses GALEX and WISE photometry, which do not have the resolving power to allow for a semi-resolved approach. However, in this subsection, we reproduce our results that rely predominantly on \emph{global} SFR and M$_\star$ measurements, now using values drawn from GSWLC. For this test, we recalculate $\Delta$SFR and $f_\text{gas, total}$ using optical products from GSWLC, the same ALMA-derived M$_\text{mol}$, and equations \ref{dsfr} and \ref{fgas}, respectively. Since GSWLC does not report stellar masses within the fibre aperture, we do not change our estimate of $f_\text{gas, inner}$, which uses M$_{\star\text{, inner}}$ in the denominator. $f_\text{gas, outer}$ uses M$_{\star\text{, inner}}$ from MPA/JHU and M$_{\star\text{, total}}$ from GSWLC.

\begin{figure*}
        \centering
        \includegraphics[width=1\linewidth]{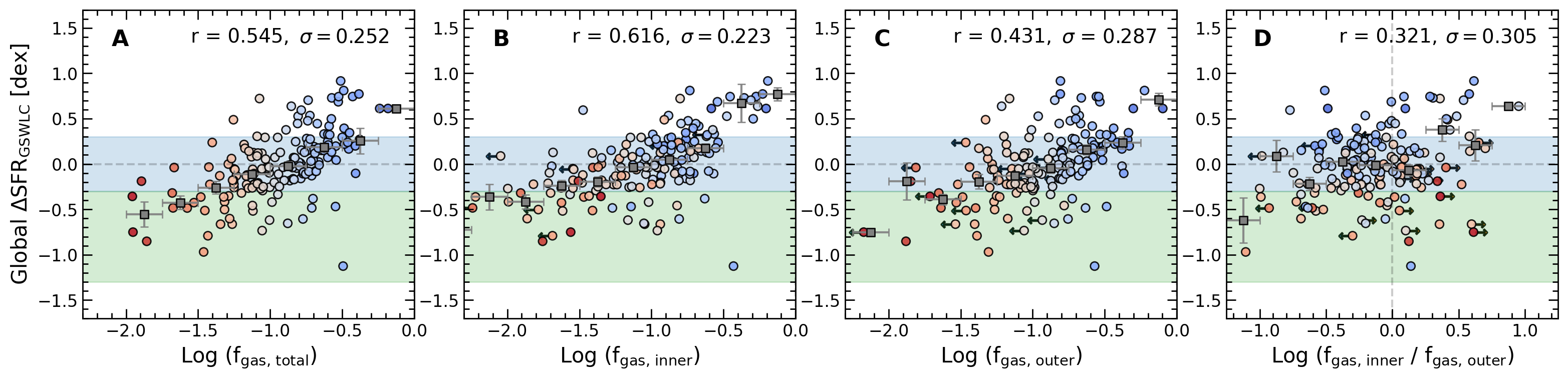}
        \caption{The same as Figure \ref{dsfr_vs_fgas}, but with global SFR and M$_{\star}$ taken from the GSWLC. Global $\Delta$SFR as a function of $f_\text{gas, total}$ (panel A), $f_\text{gas, inner}$ (panel B), $f_\text{gas, outer}$ (panel C), and the ratio between the inner and outer gas fraction (panel D). The points are the CO-detected SALVAGE galaxies coloured by their $f_\text{gas, total}$. Gray squares are the median $\Delta$SFR in bins of the different gas fraction variables. The blue and green shading represents $\Delta$SFRs consistent with star-forming and green valley galaxies, respectively. The Pearson correlation coefficient (r) and the intrinsic scatter ($\sigma$) is in the top right corner of each panel. While the intrinsic scatter is a measure of the scatter after accounting of observational uncertainty, it may not be able to account for systematic biases, if present. After changing global measurements of SFR and M$_{\star}$ our conclusion does not change: global $\Delta$SFR is most strongly correlated with $f_\text{gas, inner}$ (and not $f_\text{gas, total}$) and the relationship between $f_\text{gas, inner}$ and $\Delta$SFR has the lowest intrinsic scatter.}
        \label{dSFR_fgas_gswlc}
    \end{figure*}

Following our approach in Section \ref{results_corr}, in Figure \ref{dSFR_fgas_gswlc}, we present $\Delta$SFR$_\text{GSWLC}$ as a function of $f_\text{gas, total}$ (panel A), $f_\text{gas, inner}$ (panel B), and $f_\text{gas, outer}$ (panel C), as well as the ratio between $f_\text{gas, inner}$ and $f_\text{gas, outer}$ (panel D). In each panel, the points are coloured by $f_\text{gas, total}$ (as quantified in panel A), and non-detections in the inner or outer regions are indicated by points with arrows. The gray squares show the median $\Delta$SFR in a given gas fraction bin. For each panel, we measure the Pearson correlation coefficient and intrinsic scatter with \texttt{ltsfit}, which incorporates the uncertainty in each of the variables.

Comparing the Pearson correlation coefficients and intrinsic scatter measurements in this test with those in Figure \ref{dsfr_vs_fgas}, we find that in all cases the strength of the relationship between global $\Delta$SFR and gas fraction is weaker (in terms of both correlation and scatter) when using GSWLC measurements of SFR and stellar mass. More importantly, the relative strength of the correlations remains the same; $f_\text{gas, inner}$ correlates more strongly with $\Delta$SFR$_\text{GSWLC}$ than $f_\text{gas, total}$. This is not due to differences in the measurement of M$_{\star\text{, inner}}$ and M$_{\star\text{, total, GSWLC}}$ since they are calculated with similar methods and data (i.e. both are photometric fits), \texttt{ltsfit} accounts for different measurement errors, and the error on $f_\text{gas, inner}$ is dominated by the error on M$_\text{mol, inner}$, not M$_{\star\text{, inner}}$.

\bsp	
\label{lastpage}
\end{document}